\documentclass[aps,rmp,reprint,amsmath,amssymb,graphicx,longbibliography]{revtex4-1}

\usepackage{bm}
\usepackage{graphicx}
\usepackage{epstopdf}
\usepackage{wrapfig}
\usepackage{array} 
\usepackage{listings}
\usepackage[para,online,flushleft]{threeparttablex}
\usepackage{booktabs,dcolumn}
\usepackage{color}

\newcolumntype{C}[1]{>{\centering\arraybackslash}p{#1}}

\usepackage{textpos}
\usepackage{booktabs}
\usepackage{multirow,bigdelim}
\usepackage{float}

\usepackage{upgreek} 

\usepackage[utf8]{inputenc}
\usepackage{hyperref}
\hypersetup{breaklinks=true,colorlinks=true,linkcolor=blue,citecolor=blue,filecolor=magenta,urlcolor=blue}

\usepackage{xcolor}

\usepackage{comment}

\makeatletter
\def\@bibdataout@aps{%
\immediate\write\@bibdataout{%
@CONTROL{%
apsrev41Control%
\longbibliography@sw{%
    ,author="08",editor="1",pages="1",title="0",year="1"%
    }{%
    ,author="08",editor="1",pages="1",title="",year="1"%
    }%
  }%
}%
\if@filesw \immediate \write \@auxout {\string \citation {apsrev41Control}}\fi
}
\makeatother

\begin{document}

\title{Machine Learning in Nuclear Physics}

\author{Amber Boehnlein}
\affiliation{Thomas Jefferson National Accelerator Facility, 12000 Jefferson Avenue, Newport News, Virginia, USA}
\author{Markus Diefenthaler}
\affiliation{Thomas Jefferson National Accelerator Facility, 12000 Jefferson Avenue, Newport News, Virginia, USA}
\author{Cristiano Fanelli}
\affiliation{Laboratory  for  Nuclear  Science,  Massachusetts  Institute  of  Technology,  Cambridge,  MA  02139,  USA}
\affiliation{The NSF AI Institute for Artificial Intelligence and Fundamental Interactions}
\author{Morten Hjorth-Jensen}
\affiliation{Facility for Rare Isotope Beams and Department of Physics and Astronomy, Michigan State University, MI 48824, USA}
\affiliation{Department of Physics and Center for Computing in Science Education, University of Oslo, N-0316 Oslo, Norway}
\author{Tanja Horn}
\affiliation{Department of Physics, The Catholic University of America, N. E. Washington, DC 20064, USA}
\affiliation{Thomas Jefferson National Accelerator Facility, 12000 Jefferson Avenue, Newport News, Virginia, USA}
\author{Michelle P. Kuchera}
\affiliation{Department of Physics and Department of Mathematics \& Computer Science, Davidson College, North Carolina 28035, USA}
\author{Dean Lee}
\affiliation{Facility for Rare Isotope Beams and Department of Physics and Astronomy, Michigan State University, MI 48824, USA}
\author{Witold Nazarewicz}
\affiliation{Facility for Rare Isotope Beams and Department of Physics and Astronomy, Michigan State University, MI 48824, USA}
\author{Kostas Orginos}
\affiliation{Department of Physics, William \& Mary, Williamsburg, Virginia, USA}
\affiliation{Thomas Jefferson National Accelerator Facility, 12000 Jefferson Avenue, Newport News, Virginia, USA}
\author{Peter Ostroumov}
\affiliation{Facility for Rare Isotope Beams and Department of Physics and Astronomy, Michigan State University, MI 48824, USA}
\author{Alan Poon}
\affiliation{Nuclear Science Division, Lawrence Berkeley National Laboratory, 1 Cyclotron Road, Berkeley, California 94720, USA }
\author{Nobuo Sato}
\affiliation{Thomas Jefferson National Accelerator Facility, 12000 Jefferson Avenue, Newport News, Virginia, USA}
\author{Alexander Scheinker}
\affiliation{Accelerator Operations and Technology Division Applied Electrodynamics Group, Los Alamos National Laboratory, Los Alamos, New Mexico 87544, USA}
\author{Malachi Schram}
\affiliation{Thomas Jefferson National Accelerator Facility, 12000 Jefferson Avenue, Newport News, Virginia, USA}
\author{Michael S. Smith}
\affiliation{Physics Division, Oak Ridge National Laboratory, Oak Ridge, Tennessee, 37831-6354, USA}
\author{Xin-Nian Wang}
\affiliation{Nuclear Science Division, Lawrence Berkeley National Laboratory, 1 Cyclotron Road, Berkeley, California 94720, USA }
\author{Long-Gang Pang}
\affiliation{Key Laboratory of Quark and Lepton Physics, Institute of Particle Physics, Central China Normal University, Wuhan 430079, China}

\author{Veronique Ziegler}
\affiliation{Thomas Jefferson National Accelerator Facility, 12000 Jefferson Avenue, Newport News, Virginia, USA}

\begin{abstract}
    Advances in machine learning methods provide tools that have broad applicability in scientific research.  These techniques are being applied across the diversity of nuclear physics research topics, leading to advances that will facilitate scientific discoveries and societal applications. 
    This Colloquium provides a snapshot of nuclear physics research  which has been transformed by   machine learning techniques. 
\end{abstract}

\maketitle

\tableofcontents

\section{Introduction} 

This Colloquium represents an up-to-date summary of  work in the application of machine learning (ML)
in nuclear science, covering topics in nuclear theory, experimental methods, accelerator technology, and nuclear data.
An overview of the use of artificial intelligence (AI) techniques in nuclear physics which aimed at identifying commonalities and needs has been provided in~\citet{Bedaque2021}. 

\begin{figure*}[!htb]
\includegraphics[width=1.0\linewidth]{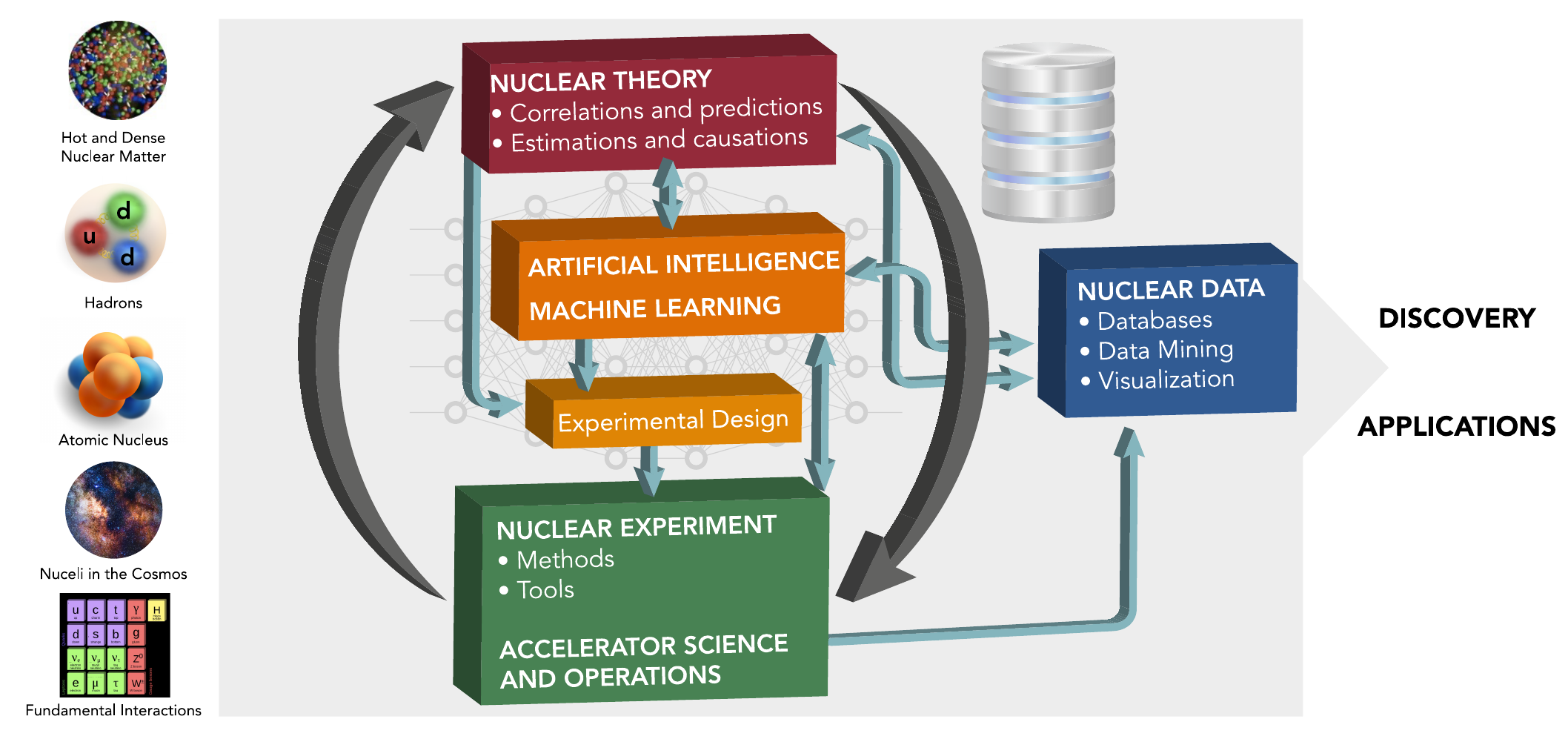}
\caption{\label{fig1} Schematic relationships between the topics discussed in this Colloquium. The diagram emphasizes the close connections between theory, computations (both computational science and data science as well as many elements from computer science) and experiments. }
\end{figure*}

Nuclear physics is a well-established field, with more than a century of fundamental discoveries covering a huge span of degrees of freedom, energy scales and length scales, ranging from our basic understanding of  fundamental constituents of matter  to the structure of stars and the synthesis of the elements in the Cosmos, see Fig.~\ref{fig1}. Experiments 
produce data volumes that range in complexity and heterogeneity, thereby posing enormous challenges to their 
 design,  their execution, and
the statistical data analysis.
 
 Theoretical modeling of nuclear properties is, in most physical cases of interest, limited by the large amount of degrees of freedom in quantum-mechanical  calculations. The analysis of experimental data and the theoretical modeling of nuclear systems aims, as is the case in all fields of physics, at uncovering the basic laws of motion in order to make predictions and estimations, as well as finding correlations and causations  for strongly interacting matter. The broad aims of nuclear physics as a field correspond to a highly distributed scientific enterprise. Experimental efforts utilize many laboratories worldwide, each with unique operation, data acquisition, and analysis methods. Similarly, the scales of focus spanned in theoretical nuclear physics lead to broad needs for algorithmic methods and uncertainty quantification. These  efforts, utilizing arrays of data types across size and energy scales,  create an ideal environment for applications of ML methods.

\section{Machine learning for nuclear physics in broad strokes}

 Statistics, data science, and ML form important fields of research in modern science. They describe how to 
 learn and make predictions from data, and enable the extraction of key information about physical processes and the underlying scientific laws based on large datasets. As such, recent advances in ML capabilities are being applied to advance scientific discoveries in the physical sciences \cite{deiana2021,Carleo2019}.

Ideally, ML represents the science of building models to perform a task without the instructions being explicitly programmed. This approach introduces in practice a hierarchy of mathematical operations that enable the computer to learn complicated concepts by building them out of simpler ones. In terms of a graphical representation, this can be visualized as a deep network of training and learning operations, often just referred to as deep learning \cite{Goodfellow2016}. 

There exist  many ML approaches; they are often split into two main categories, supervised and unsupervised. In supervised learning, training data are labeled  and one lets a specific ML algorithm learn and deduce patterns in the datasets. 

This allows one  to make predictions about future events and/or data not included in the training set.
On the other hand, unsupervised learning is a method for finding patterns and relationship in datasets without any prior knowledge of the system. Many researchers also operate with a third category, dubbed reinforcement learning. This is a paradigm of learning inspired by behavioral psychology, where actions are learned to maximize reward.
One may  encounter reinforcement learning being accompanied by supervised deep learning methods such as deep ANN. Furthermore, what is often referred to as  semi-supervised learning, entails developing  algorithms that aim at learning from a dataset that includes both labeled and unlabeled data.

Another way to categorize ML tasks is to consider the desired output of a system. Some of the most common tasks are \cite{Hastie2009,Murphy2012,Goodfellow2016,Bishop2006,Mehta2019,Cranmer2021}:
\makeatletter
\renewenvironment{description}%
               {\list{}{\leftmargin=10pt 
                        \labelwidth\z@ \itemindent-\leftmargin
                        \let\makelabel\descriptionlabel}}%
               {\endlist}
\makeatother
\begin{description}
    \item[Classification] Outputs are divided into two or more classes. The goal is to produce a model that assigns inputs into one of these classes. An example is to identify digits based on pictures of hand-written numbers. 
\item[Regression] Finding a functional relationship between an input dataset and a reference dataset. The goal is to construct a function that maps input data into continuous output values.
\item[Clustering] Data are divided into groups with certain common traits, without knowing the different groups beforehand. This ML task falls  under the category of unsupervised learning.
\item[Generation] Building a model to generate data that are akin to a training dataset in both examples and distributions of examples. Most generative models are types of unsupervised learning.
\end{description}
In Table\,\ref{tab:acronyms} we list many of the methods encountered in this work, with their respective abbreviations.
\begin{table*}[!htb]
 \caption{Table of ML methods discussed in this Colloquium  with indication of the main type of learning (S: supervised,  U: unsupervised,  Semi-S: semi-supervised).
 }
    \label{tab:acronyms}
\begin{ruledtabular}
    \begin{tabular}{C{0.05\linewidth} p{0.25\linewidth}
    p{0.40\linewidth}C{0.12\linewidth}}
    Acronym & 
  \multicolumn{1}{c}{Method} & 
      \multicolumn{1}{c}{Brief Description} & Learning Type\\ \hline\\[-5pt] 
AE/VAE & Auto Encoders / Variational Auto Encoders  & ANN capable of learning efficient representations of the input data without
any supervision  & U \\
    ANN     & Artificial Neural Networks & Models for learning defined by  connected units (or nodes) and hidden layers with well defined inputs and outputs.  & S  \\ 
  BED & Bayesian Experimental Design   & Bayesian inference for  experimental design  & S \\
        BM & Boltzmann Machines  & Generative ANN that can learn a probability distribution from sets of changing inputs  & U \\
    BMA/BMM & Bayesian Model Averaging/ Mixing  & Bayesian inference applied to model selection, combined estimation, or  performed over a mixture model & S \\
        BO & Bayesian Optimization  & Optimization of functions  without an {\em a priori}  knowledge of functional forms. & S and Semi-S\\
    BNN & Bayesian Neural Networks  & ANN where the parameters of the network are represented by probabilities learnt by Bayesian inference & S \\
    CNN & Convolutional Neural Networks  & ANN where convolution is used to reduce dimensionalities & S \\
     EMB  & Ensemble Methods  {\&} Boosting
      & Methods based on collections of decision trees as simple learners & S \\
    GAN & Generative Adversarial Networks   & System of two ANN where a generative network generates outputs while a discriminative network evaluates them & U \\
    GP & Gaussian Processes   & Collection of random variables which have a joint Gaussian distribution used in Bayesian inference & Semi-S \\
    KNN & $k$-nearest neighbors  &  Non-parametric method where inputs consist of the $k$ closest training examples in a dataset & S \\
    KR & Kernel Regression  & Extension of linear regression methods to include non-linear function kernels & S \\
    LR & Logistic Regression  & Convex optimization method based on maximum likelihood estimate for classification problems & S\\ 
    LSTM & Long short-term memory  & RNN capable of learning long-term dependencies & S\\
    PCA & Principal Component Analysis 
     & Dimensionality reduction technique based on retaining the largest eigenvalues of the covariance matrix & U \\
    REG    & Linear Regression   & Linear algebra methods used for modeling continuous functions in terms of their explanatory variables & S \\
    RL & Reinforcement Learning  & Learning  achieved by trial-and-error of desired and undesired events & Neither S nor U \\
    RNN & Recurrent Neural Networks  & ANN where connections between nodes allow for temporal dynamic behavior & S \\
    SVM & Support Vector Machines  & Convex optimization techniques with efficient ways to distinguish features in datasets & S \\
     \end{tabular}
 \end{ruledtabular}  
\end{table*}

The methods we cover here have three 
central elements in common, irrespective of whether we deal with supervised, unsupervised, or semi-supervised learning. The first element is some  dataset (which can be subdivided into training, validation,  and test data), the second 
element is a model, which is normally a function of some parameters to be determined by the chosen optimization process. The model reflects our prior knowledge of the system (or lack thereof). As an example, if we know that our data show a behavior similar to what would be predicted by a polynomial, fitting the data to a polynomial of some degree would  determine our model.
The last 
element is a so-called cost (or loss, error, penalty, or risk) function which allows us to present an estimate on how good our model is in reproducing the data it is supposed to train. This is the function which is optimized in order to obtain the best prediction for the data under study. The simplest cost function in a regression analysis (fitting a continuous function to the data) is the so-called mean squared error function 
while for a binary classification problem, the so-called cross entropy is widely used, see for example Refs.~\cite{Goodfellow2016,Murphy2012,Bishop2006,Hastie2009}
for more details. We will henceforth refer to this element as the assessment of a given method.

Traditionally, the field of ML has had its main focus on predictions and correlations.
In ML and prediction-based tasks, we are often interested in developing algorithms that are capable of learning patterns from 
existing data in an automated fashion, and then using these learned patterns to make predictions or assessments of new  data. In some cases, our primary concern is the quality of the predictions or assessments, with perhaps less focus on the underlying patterns (and probability distributions) that were learned in order to make these predictions.  However, in many  nuclear physics studies, we are equally interested in being able to estimate errors and find causations.  
In this Colloquium, we emphasize the role of predictions and correlations as well as error estimation and causations in statistical learning and ML.  For general references on these topics and discussions of frequentist and Bayesian methodologies, see, e.g., \cite{Bishop2006, Goodfellow2016, Murphy2012, Hastie2009, Myren2021,Gelman2014}. 

Since the aim of this Colloquium is to give an overview of present usage and research of ML in nuclear physics, we utilize material from several excellent sources on the topic, such as the textbooks mentioned above  and recent  reviews  \cite{Mehta2019,Carleo2019}. We would also like to point to the theory of Bayesian Experimental Design (BED) \cite{Chaloner1995,Liepe2013} -- a theory that is tailored for making optimal decisions under uncertainty.

During the last few years there has been a surge of interest in applying different ML and Bayesian methods in nuclear physics. In particular, a Bayesian approach has gained large traction since an estimation of errors plays a major role in theoretical  studies, such as reliable determinations of parameters entering models for nuclear forces and density functionals. Similarly, in quantum-mechanical few- and many-body studies, a number of research groups have implemented ML-based techniques in order to handle complicated correlations and exploding numbers of degrees of freedom. These studies cover a large set of approaches, ranging from applications of deep learning methods, such as ANN and restricted BM for solving the many-particle Schr{\"o}dinger equation,  to ANN in many-body methods, with the aim to learn many-body correlations. Similarly, there have been several attempts to use ML approaches to
extract information about correlations in field theory and lattice quantum chromodynamics (LQCD), ranging from attempts to circumvent the fermion sign problem to learning fermion propagators and more. For recent ML applications to nuclear theory, see Sec.~\ref{Theory}.

In experimental nuclear physics, with increasing degrees of freedom and complexity, one faces many of the same challenges as in nuclear theory. As discussed in Sec.~\ref{Experiment}, ML approaches offer a number of  optimization strategies to handle this surge in dimensionality.
Many nuclear physics experiments nowadays produce huge amounts of data in excess of terabytes, requiring the use of fast algorithms for tractable data collection and analysis. Machine learning methods such as anomaly detection allow for exploration of data  for unforeseen phenomena. Additionally, labeled data may not be available, either due to the inability to label data or lack of knowledge of the types and behaviors of the reactions taking place. The latter are normally needed in order to generate simulated data one can use in the training process. Machine learning techniques play also a considerable role in accelerator science and operations and nuclear data science, see Sections~\ref{Accelerators} and \ref{Data}, respectively.

After these rather general remarks about ML in nuclear physics, we move on to a description of ongoing and planned research where many of these approaches  are applied to  multidimensional problems, large datasets, detection and prognostication, design optimization, and real time operational control.

\section{Nuclear Theory}\label{Theory}

The aim of this section is to give the reader an overview of recent progress and future research directions in ML approaches and methods applied to nuclear theory. During the last few years, ML methods have been applied to essentially all length and energy scales of interest for nuclear theory, spanning from theories for the strong force to the equation of state for  neutron stars. We start our discussion with low-energy nuclear theory, moving up to medium energies,  and to high-energy nuclear theory and lattice quantum chromodynamics.

\subsection{Low-Energy Nuclear Theory}\label{LENT}

\subsubsection{Early applications of machine learning} 
In a pioneering paper, the St.Louis-Urbana collaboration \cite{Gazula1992} successfully carried out computer experiments based on ANN to study global nuclear properties across the nuclear landscape, including dripline locations, 
atomic masses, separation energies, and location of shell-stabilized superheavy nuclei. Their work  did recognize the potential of using ML techniques to describe the variety of nuclear behavior: ``{\it The field of nuclear physics, with a wealth of data reflecting both the fundamental principles of quantum mechanics and the behavior of strong, electromagnetic, and weak interactions on the Fermi scale of distances, offers especially fertile ground  for testing and exploiting the new concept of adaptive phenomenological analysis based on neural networks.}" They concluded that ``{\it Significant predictive ability is demonstrated, opening the prospect that neural networks may provide a valuable new tool for computing nuclear properties and, more broadly, for phenomenological description of complex many-body systems.}" Encouraged by the ability of the ANN to capture the patterns and irregularities of nuclear observables, they extended their investigations to 
systematics of  nuclear spins and parities \cite{Gernoth1993}. The SVM  study of  nuclear masses, beta-decay lifetimes, and spins/parities of nuclear ground states was reported in \cite{Clark:2006ua,Costiris2008}, and the application of ANN  to beta-decays was carried out in \cite{Costiris:2008pp}.

\subsubsection{Machine learning for data mining}
Oftentimes, it is necessary to be able to accurately calculate observables that have not been measured, to supplement the existing databases. To provide quantified  interpolations and extrapolations of nuclear data, nuclear models  augmented by modern ML techniques have been used. Examples include: studies of nuclear masses with EMB \cite{Carnini2020}, KR \cite{Wu2020a,Wu2021}, and ANN \cite{Yuksel2021,Lovell2022}; calculations of the nuclear charge radii using ANN \cite{Wu2020}; estimating of $\alpha$-decay rates using EMB and ANN \cite{Saxena2021}; constraining fission yields using mixture-density ANN \cite{Lovell2019,Lovell2020}, BNN \cite{Wang2019,Qiao2021,wang2021bayesian,Wangpei2021}, and KNN \cite{TongL2021};
  estimation of the  total fusion cross-sections using ANN \cite{Akkoyun2020}; prediction of the isotopic cross-sections in proton-induced spallation reactions using BNN \cite{Ma2020}; and estimation of gamma strength functions using BO \cite{Heim2020}.

\begin{figure}[htb]
\includegraphics[width=0.8\linewidth]{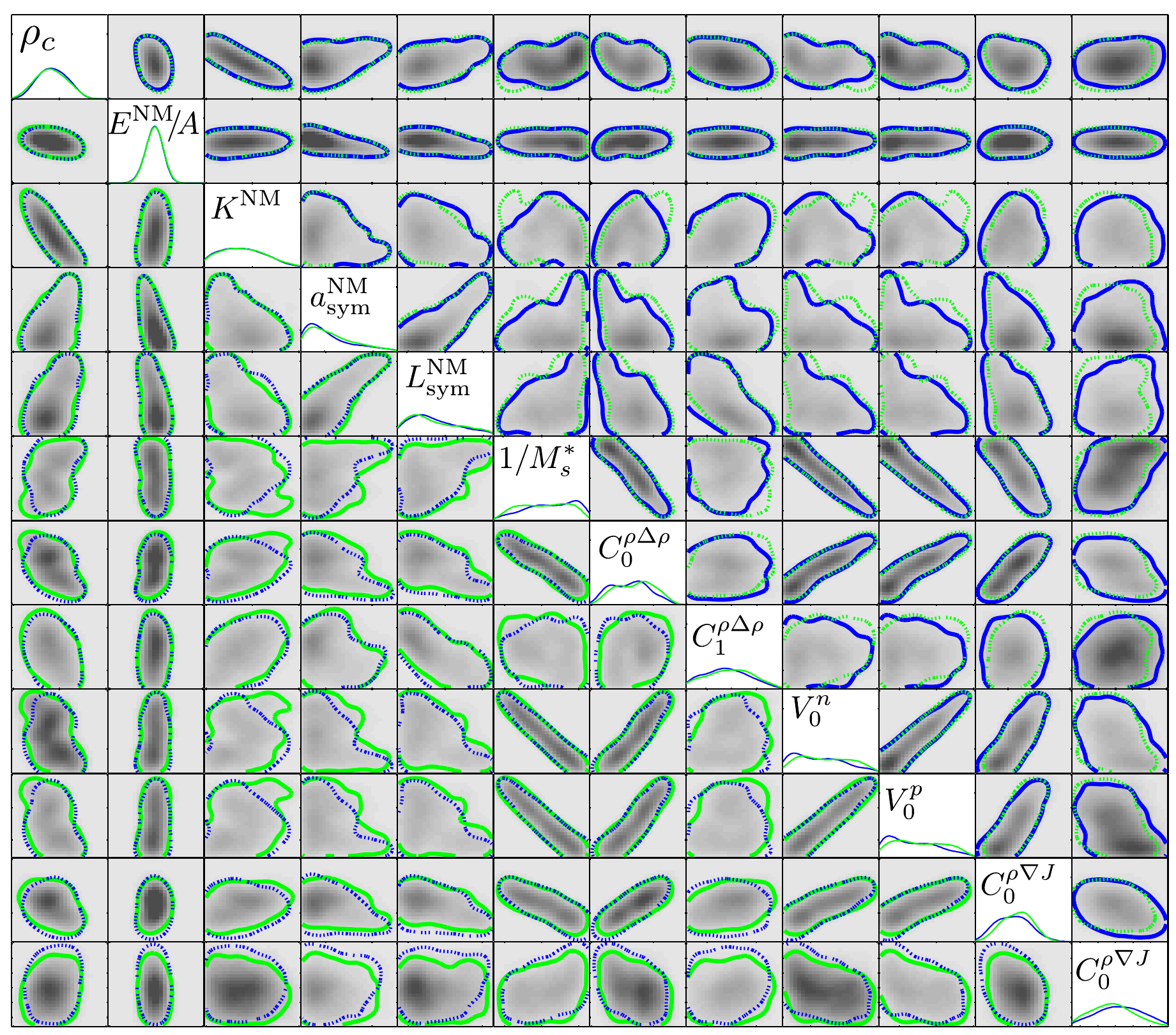}
\caption{\label{DFT-Higdon} {\bf Bayesian calibration of energy density functionals.} Univariate and bivariate marginal estimates of the posterior distribution for the 12-dimensional parameter vector of the UNEDF1 EDF. The blue lines enclose an estimated 95\% region for the posterior distribution found when only the original UNEDF1 data are accounted for; the green-outlined regions represent the same region for the posterior distribution found when the more recent  mass measurements are included. Taken from \cite{McDonnell2015}.}
\end{figure}

\begin{figure}[htb]
\includegraphics[width=1.0\linewidth]{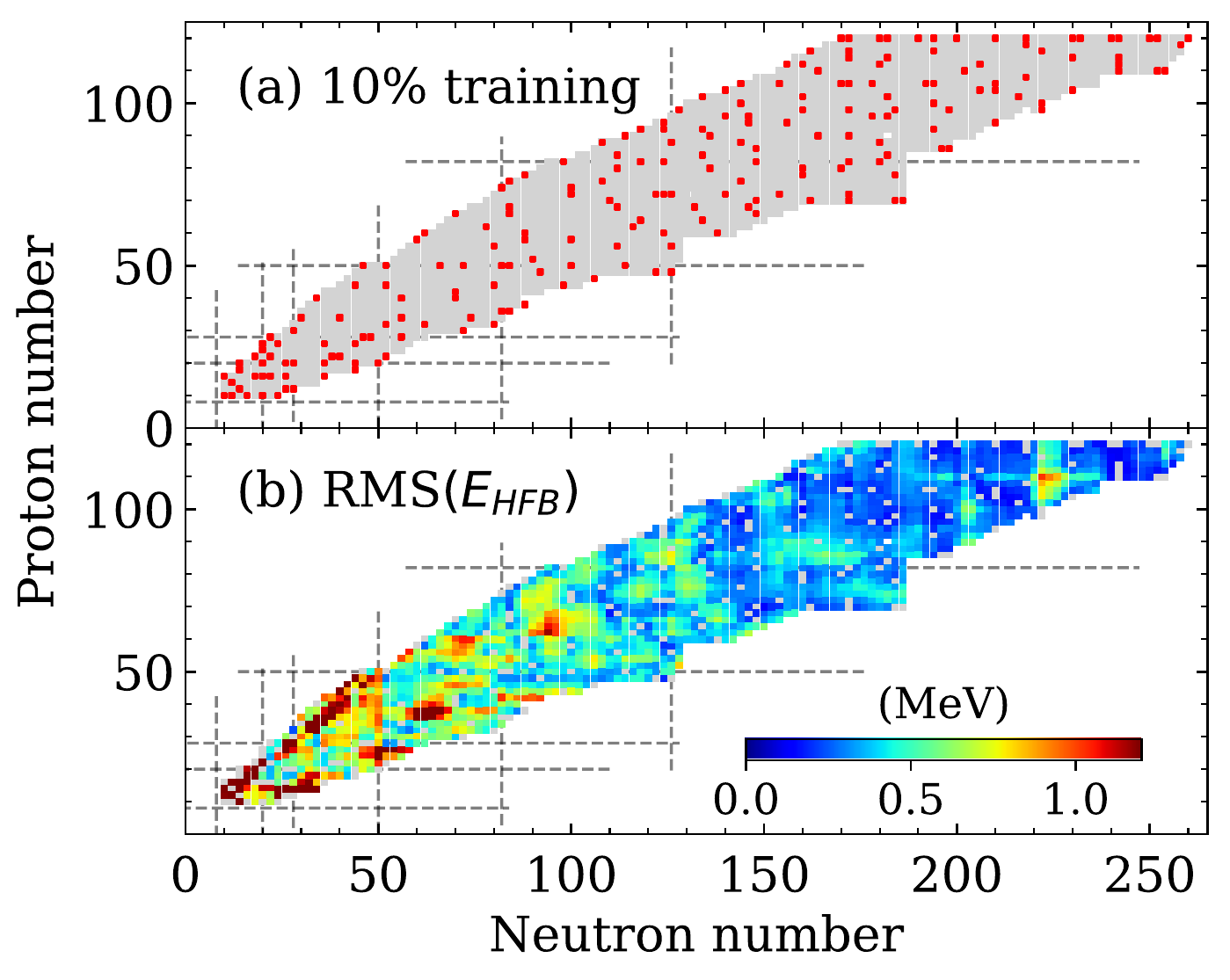}
\caption{\label{DFT-EPS} {\bf DFT emulator with ANN.} (a) The database nuclei  (grey) as a function of $N$ and $Z$.  Nuclei included in the 10\% training dataset obtained by the active learning are marked in red. (b) The root mean square deviation between the total energy for the testing dataset calculated in DFT ($E_{HFB}$)  and with the committee of ANN. Adopted from \cite{Lasseri2020}.}
\end{figure}

\subsubsection{Properties of heavy nuclei and nuclear density functional theory}
Kohn–Sham density functional theory (DFT) \cite{KohnSham1965}  is the basic computational approach to multi-electron systems and there exists a rich literature on ML applications in the field of the  electron DFT \cite{Carleo2019,Hautier2010,Schleder2019,Ryczko2019,Nagai2020,Moreno2020}. Nuclear DFT,  rooted in the self-consistent  mean-field approach \cite{Bender2003,Schunck2019B,Yang2020b}, is the basic computational framework for
the global modeling of all nuclei, including complex exotic nuclei far from stability. 
An effective interaction in DFT is given by the energy density functional (EDF),
whose parameters are adjusted to experimental data.
Over the past decade, better and more refined EDFs  have been developed, with increasingly complex and computationally expensive computer models, often involving BO and ANN ML  \cite{McDonnell2015,Higdon2015,Goriely2014,Navarro2018,Kejzlar2020,Scamps2020,Bollapragada2020,Schunck2020,Zhang2021}. Figure \ref{DFT-Higdon} shows the  posterior distributions for the  parameters of the UNEDF1 EDF obtained in \cite{McDonnell2015} by means of the Bayesian model calibration. These distributions have been used to provide uncertainty quantification (UQ) on UNEDF1 model predictions, in particular  the
$r$-process abundance pattern \cite{Sprouse2020}, 
 nuclear matter equation of state \cite{Du2019},
 and neutron star properties \cite{Almamun2021}.

Since global DFT computations are expensive, a promising avenue for ML applications is the emulation of DFT results \cite{Akkoyun2013,Lasseri2020,Schunck2020a,Apurba2021}.  
Figure~\ref{DFT-EPS} shows the results of ANN calculations of \cite{Lasseri2020}: a committee of ANN
trained on a  set of 210 nuclei is capable of predicting the ground-state and excited energies of more than 1800 atomic nuclei with significantly less computational cost than full DFT calculations. 

\begin{figure*}[!htb]
\includegraphics[width=0.7\linewidth]{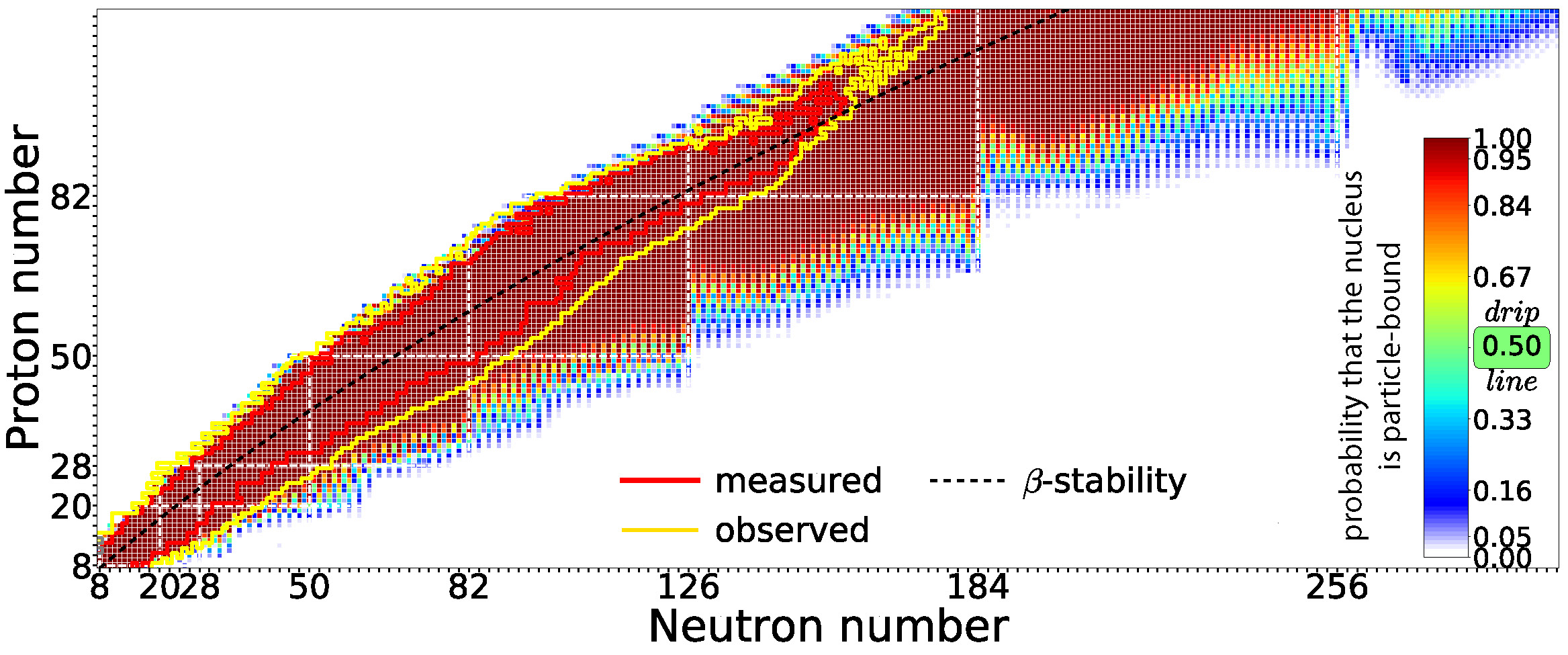}
\caption{\label{BMA} {\bf Bayesian extrapolation and model averaging.} The quantified  landscape of nuclear existence obtained in the BMA calculations using 11 global mass models and three model-averaging strategies. For every nucleus with $Z,N \ge 8$ and $Z\le 119$ the probability that  the nucleus is bound with respect to proton and neutron decay, is marked. The domains of nuclei which have been experimentally observed and whose separation energies have been  measured (and  used for training) are indicated.  Adopted from \cite{Neufcourt2020a}.}
\end{figure*}
\subsubsection{Nuclear properties with ML}
One can improve the predictive power of a given nuclear model by comparing its predictions to existing data. Here, a powerful strategy
is to estimate model residuals, i.e., deviations between experimental and calculated observables, by developing an emulator  using a training set of observables taken from experiment or other theory. An emulator can be constructed by employing Bayesian
approaches, such as BNN and GP. Global surveys of nuclear observables employing such a strategy can be found in
\cite{Utama16,Utama17,Utama2018,Niu2018,Niu2019,Rodriguez2019,Pastore2020,Ma2020a,Gao2021} (extrapolations of nuclear masses with NN);
\cite{Neufcourt2018,Shelley2021} (extrapolations of nuclear masses with GP);
\cite{Utama16a,Wu2020} (studies of nuclear radii);
\cite{Niu2018a,Wu2021_beta} (studies of beta decay rates); and
\cite{Rodriguez2019a} (studies of alpha decay rates). See also \cite{Pastore2021,Liu2021} for more discussion on ANN extrapolations.

By considering several global models and the most recent
data one can apply the powerful techniques of Bayesian model averaging (BMA) and Bayesian model
mixing (BMM) to assess model-related  uncertainties  in the multi-model context \cite{Phillips2021}.
Examples of recent model-mixed predictions using BMA include:
analysis of the neutron dripline in the Ca region \cite{Neufcourt2019a}; studies of proton dripline and proton radioactivity \cite{Neufcourt2020}; quantifying the particle stability of nuclei \cite{Neufcourt2020a};
 combining models calibrated in different domains \cite{Kejzlar2020}; and assessing the puzzling mass of $^{80}$Zr \cite{Hamaker2021}. Figure~\ref{BMA}  shows the posterior probability of existence  for all nuclei in the nuclear landscape based on predictions of eleven global mass models, the most recent data on nuclear existence and masses, and three model-averaging strategies  to compute the BMA weights.

\subsubsection{Nuclear shell model applications} Machine Learning methods have been used to provide UQ of shell model (configuration interaction) calculations based on phenomenological two-body matrix elements. 
In \cite{Yoshida2018}, Bayesian ML was used to
compute marginal estimates of the posterior distribution for the  shell-model Hamiltonian in the $0p$ space
and uncertainty estimates on observables. A similar analysis,
but for effective Hamiltonians in the $1s0d$-shell,
was carried out in  \cite{Fox2020,Magilligan2020} (using PCA) and \cite{Akkoyun2020a} (using ANN).
Eigenvector continuation (EC) was used in 
\cite{Yoshida2021} to construct  an emulator of the shell-model calculations for a valence space, parameter optimization, and  UQ.

\subsubsection{Effective field theory and $A$-body systems}
Bayesian ML, BO, and UQ in {\em ab initio} nuclear theory are reviewed in \cite{Ekstrom2019,Ekstrom2019b}.
There have been several studies of nucleon-nucleon scattering using Bayesian ML to estimate chiral effective field theory (EFT)  truncation errors and low-energy coupling constants \cite{Melendez2017,Wesolowski2018,Melendez2019,Svensson:2021lzs}, and there have also been applications to $np \rightarrow d\gamma$ \cite{Acharya:2021lrv}. In Ref.~\cite{Furnstahl2020}, the authors used  eigenvector continuation \cite{frame2018} as an emulator for scattering. The method has been further improved in \cite{Drischler:2021qoy} by making use of different boundary conditions. The application of EC as a fast emulator for UQ for few- and many-body systems was first explored by \cite{Konig2020} and adapted for coupled cluster calculations and a global sensitivity analysis of $^{16}$O \cite{Ekstrom2019a} and, with Bayesian history matching \cite{Vernon2014}, to global properties of  $^{208}$Pb \cite{hu2021ab}.   Eigenvector continuation emulators have been used to put rigorous constraints on low-energy constants for the three-nucleon forces \cite{Wesolowski2021} and make predictions for binding of $A=6$ nuclei \cite{Djarv:2021ckm}.   For other applications of eigenvector continuation, see \cite{Melendez2021a,Eklind:2021car,Zhang:2021jmi,Sarkar:2021fpz}.  In Ref.~\cite{Connell2021}, the authors address whether BMA improves the
extrapolation of polynomials, which are used as proxies for fixed-order EFT calculations.

Using a scattering amplitude as input data (respecting unitarity, hermiticity and analyticity), 
\cite{Sombillo2020,Sombillo2021,Sombillo:2021ifs} classified  the nature of the poles near threshold  with  multi-layer ANN. In particular, in \cite{Sombillo2020} the ANN were applied to predict the correct nature of the poles in the nucleon-nucleon scattering data from a partial wave dataset. This is an example of a classification problem where the aim is to classify whether the poles represent bound, virtual, or resonant states.  \cite{Kaspschak:2020zws} proposed   an iterative ANN perturbation theory to study $s$-wave scattering lengths for shallow potentials. 

\cite{Navarro2015} propagated the statistical errors in nucleon-nucleon scattering to calculations of light nuclei. The radiative capture rates $^7$Be+$p\rightarrow$ $^8$B+$\gamma$ \cite{Zhang:2015ajn} and $^3$He+$^4$He$\rightarrow$ $^7$Be+$\gamma$ \cite{Zhang2020} were estimated with BO.  Bayesian ML has been applied to neutron-$\alpha$ scattering in chiral EFT \cite{Kravvaris2020}.

\begin{figure}[htb]
    \includegraphics[width=1.0\linewidth]{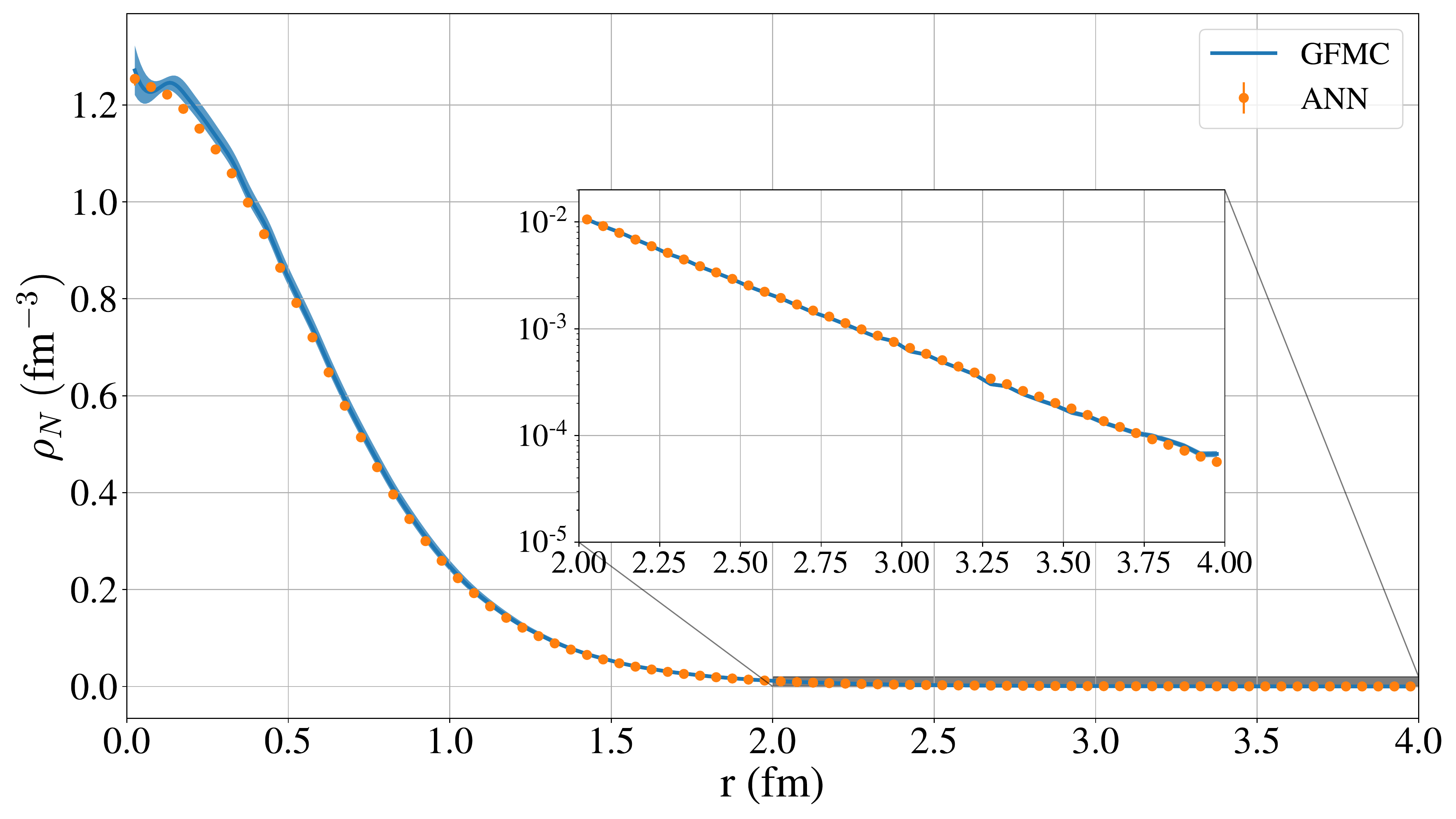}
    \caption{\label{GFMC} {\bf Many-body variational calculations with ANN.} Point-nucleon densities of $^4$He for the LO pionless-EFT Hamiltonian.  The solid points and the shaded area represent the ANN and Green's function Monte Carlo results, respectively. Taken from \cite{Adams2021}.}
\end{figure}

In other few-body calculations, ANN  were used to determine the deuteron wave function using variational optimization \cite{Keeble2020}.   Three-body Efimov bound states were studied using ANN \cite{Saito2018}, and  CNNs were used to classify states of a  three-body system \cite{Huber:2021fau}.  Variational Monte Carlo calculations optimized with ANN have been performed for light nuclei with up to six nucleons \cite{Adams2021,Gnech2021}, see Fig.~\ref{GFMC}. The latter results are  interesting since the representation of the Jastrow factor  in terms of  ANN has the potential to introduce additional correlations. The universal approximation theorem \cite{Cybenko1989,Hornik1991} states that ANN  can represent a wide variety of non-linear functions when given appropriate weights. Overall, using ANN to extract complicated correlations in many-body calculations seems to be a very promising approach, as shown in the recent work by \cite{pescia2021}. Therein, ANN were used  for the simulation of strongly interacting systems in the presence of spatial periodicity. This has potential applications  for studies of infinite nuclear matter, where periodic boundary conditions are often imposed to extract the equation of state of dense fermionic matter, see discussion below.

In larger $A$-body systems, the ANN extrapolation of nuclear structure observables have been used for the no-core shell model \cite{Negoita2019}, coupled cluster theory \cite{Jiang2019}, and configuration interaction calculations \cite{Yoshida2020}.  Artificial neural networks were used to learn important configurations for symmetry-adapted no-core shell-model calculations  \cite{Molchanov2021}. Similarly,
ANN have also been used to invert Laplace transforms required to compute nuclear response functions from Euclidean time Monte Carlo data \cite{Raghavan2021}.  Restricted BM were used to represent many-body contact interactions using auxiliary fields \cite{Rrapaj2020}.  \cite{Ismail:2021} used ML techniques to study finite-size effects and extrapolate the unitary gas to the thermodynamic limit at zero range.

Another topic which has a great potential for applications to studies of infinite matter and the equation of state (EoS) for dense nuclear matter, is the application of ML to  many-body methods like coupled cluster theory, Green's function theories, in-medium similarity renormalization group methods,  see, e.g.,  \cite{Hjorthjensen2017}. 
Common to these methods is that the underlying approximations  can be systematically expanded upon by including more complicated correlations. The various approaches, like for example   coupled cluster theory \cite{Hjorthjensen2017}, sum  to infinite order selected many-body contributions such as
so-called one-particle-one-hole and two-particle-two-hole correlations.   Including three-particle-three-hole correlations is computationally much more involved. Here, ML-based methods can be extremely useful, in particular for studies of nuclei from calcium and beyond and infinite matter. The recent atomic and molecular physics studies \cite{Margraf2018,Wilkins2019,Agarwal2021,Townsend2020}  can easily be extended to  finite nuclei and infinite matter.

\subsubsection{Nuclear reactions}
Low-energy nuclear reaction models have a critical reliance on a wide variety of parameters including nuclear masses, nuclear level densities, transmission coefficients, and optical model parameters. While nuclear masses were discussed above, there are some ML-based studies seeking improved predictions of other parameters across the nuclear chart, see Sec.~\ref{Data}. There are also numerous studies using BO for UQ on reaction model parameters; some recent examples include $R$-matrix analyses of cross sections \cite{Odell:2021tqd}, optical model parameters \cite{Lovell2018,King2019,Lovell2020a,Yang2020}, and sensitivity analyses \cite{Catacora-Rios2019,Catacorarios2020}. In the future, it is anticipated that ML will help identify those measurements that most effectively constrain theoretical models, optimize model parameters simultaneously across multiple reaction channels for many isotopes, and provide guidance to theory through global systematic studies that can be efficiently executed with surrogate models.

\subsubsection{Neutron star properties and nuclear matter equation of state}
Studies of dense nuclear matter and its pertinent EoS, with its strong implications for studies of neutron stars and studies of supernovae, is a field which has seen a considerable progress during the last two decades. The increased wealth of data related to cold dense matter, from laboratory experiments and theoretical simulations  to neutron star observations such as the gravitational-wave events GW170817 and GW190814,  provide a framework for constraining theoretically the EoS of dense matter. 
\begin{figure}[hbtp]
  \includegraphics[width=1.0\linewidth]{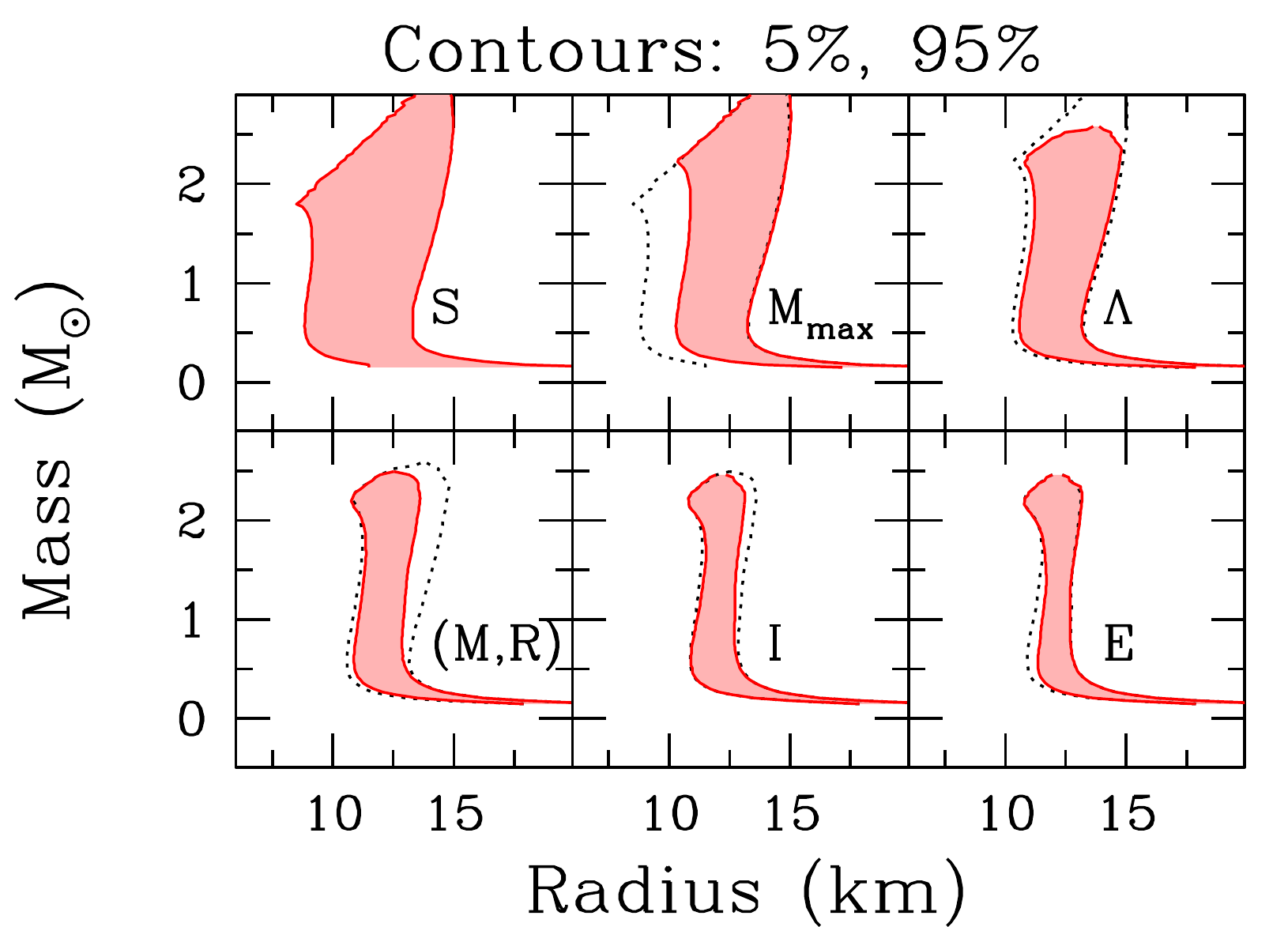}
   \caption{{\bf Bayesian analysis of the mass-radius relation for neutron stars}. The Bayesian uncertainty reflects constraints on  an equation of state.  Here $S$ is the symmetry energy; $M_{\rm max}$ - masses of three most massive neutron stars; $\Lambda$ - tidal deformability of GW170817; $(M,R)$  -  high-precision mass-radius measurement;
 $I$ - moment of inertia; and $E$ - baryonic rest mass of a star.
 Taken from \cite{Miller2019}.}
\label{fig:FigMiller2019}
\end{figure}

As an example, we present in Fig.~\ref{fig:FigMiller2019} the results of a Bayesian analysis by \cite{Miller2019} based on a posterior probability distribution. 
The figure displays the mass-radius constraints which correspond to an EoS obtained from  Bayesian inference. Here a symmetry energy of $S=32\pm 2$ MeV was adopted, together with the masses of the three most massive neutron stars, the tidal deformability of GW170817,  hypothetical masses and radii to a precision of $5\%$, and similarly a hypothetical measurement of the moment of inertia of a $1.338M_{\odot}$ star to $10\%$ precision.  The authors show that the resulting EoS is sensitive to the symmetry energy below saturation density. 

In several studies,  Bayesian inference, ML,  and/or other statistical approaches, have been applied  to  constrain the EoS and other properties pertaining to infinite matter studies. For example, \cite{Drischler2020,Drischler2020a} carried out UQ of many-body calculations of the EoS. In particular,  Bayesian ML with GP were employed to propagate theoretical uncertainties using many-body perturbation theory to fourth order with two- and three-body interactions from chiral effective field theory. Gaussian processes were also used in  \cite{Essick2021} to constrain the symmetry energy and its  slope. 
Deep learning inference for the EoS was recently studied  in \cite{Fujimoto:2021zas} using observational data on masses and radii of known neutron stars.  Several other groups have used ML methods and/or Bayesian inference to study the nuclear EoS with nuclear data and theoretical calculations \cite{Margueron2018,Margueron2018a,Xu2020,Newton2021}, X-ray observations of neutron stars \cite{Nattila2017}, gravitational wave data from neutron star mergers
\cite{Lim2019,Capano2020,Dietrich2020,Guven2020,Kunert2021}, or both X-ray and gravitational wave data \cite{Xie2020,Raaijmakers2020,Han2021,Almamun2021,Ayriyan2021}.  The properties of dense matter have also been probed using heavy-ion collisions \cite{Morfouace2019,Xie2020a} as well as studies at nonzero temperature \cite{Wang2020}.  At intermediate energies, CNN have recently been applied in \cite{zhang2021determine} to determine the impact parameters of heavy-ion collisions at low to intermediate incident energies (up to one hundred MeV/nucleon).

\subsection{Medium-Energy Nuclear Theory}

Nuclear femtography is an emerging field in nuclear physics that aims to map out quantum correlation functions (QCFs) that characterize the internal three-dimensional structure of nucleons and nuclei, as well as hadronization in high-energy reactions, in terms of the quark and gluon (collectively called partons) degrees of freedom of quantum chromodynamics (QCD). In contrast to any system found in nature, the partons of QCD are not detectable experimentally due to confinement, which prevents direct access to the QCFs. Fortunately, certain classes of observables can be factorized in terms of convolutions between short-distance physics calculable in perturbative QCD and long-distance physics which is non-perturbative and encoded in the formulation of QCFs. However, in order to extract the latter, one faces an inverse problem inherited from the mathematical relation between the experimental observables and the QCFs stemming from the inability to obtain closed form solutions for the QCFs. Therefore, the only practical approach is to parameterize the QCFs and calibrate them via, e.g.,  example BO.  

\subsubsection{Bayesian inference}
The traditional approach to implementing Bayesian inference involves the use of theory-inspired parameterizations for the QCFs, tuned via maximum likelihood estimators \cite{Hastie2009,Murphy2012,Bishop2006}, along with an error analysis based on the Hessian matrix optimization \cite{Murphy2012,Bishop2006,Pumplin:2002vw}. This approach was developed in the context of QCD global analysis of parton distribution functions (PDFs), which is a type of QCF describing the longitudinal momentum fractions of partons inside nucleons. This approach has been adopted by various groups around the world   \cite{Harland-Lang:2014zoa,Hou:2019efy,Accardi:2016qay,Alekhin:2018pai} and has found relatively good success in describing a large bank of high-energy data that are sensitive only to the one-dimensional degrees of freedom in the nucleon. 

In recent years, Monte Carlo based methodologies for Bayesian inference \cite{Murphy2012,Hastie2009,Mehta2019} have gained traction, providing a more reliable uncertainty quantification for QCFs in the Bayesian framework. The  work of \cite{Ball:2010de} demonstrated the feasibility of carrying out QCD global analysis by sampling the Bayesian posterior distribution using the data resampling technique. They have also introduced ANN parameterizations to extend the flexibility for the QCF modeling and explore the degree of parameterization bias. For the case of PDFs, one found that in regions where the data maximally constrain the PDFs, ANN and  traditional parameterizations give qualitatively similar results, with differences becoming increasingly evident in extrapolation regions. 

\subsubsection{Simultaneus extraction paradigm}
The Monte Carlo approach for Bayesian inference has also recently been applied to a  simultaneous extraction of a variety of QCFs, including spin-dependent PDFs \cite{Ethier:2017zbq}, transverse momentum distributions \cite{Cammarota:2020qcw}, as well as fragmentation functions \cite{Ethier:2017zbq, Sato:2019yez, Moffat:2021dji}, establishing a new paradigm in nuclear femtography. The simultaneous approach is crucial for solving the inverse problem for the QCFs, especially for those quantities that rely on each other. An example of such a situation is in semi-inclusive deep-inelastic scattering, where the experimental observables are sensitive to QCFs describing the internal structure of hadrons as well as QCFs that describe the hadronization process. In principle, each of these types of QCFs can be extracted independently from processes that are solely sensitive to each type of QCFs. However, when included within a global context, they require simultaneous analysis in order to take into account the correlations induced by uncertainties on the various interdependent QCFs themselves.

\cite{Ethier:2017zbq}  showed that the strange polarization in the nucleon is mostly unconstrained if one takes fully into account the uncertainties on the spin-dependent PDFs and fragmentation functions. In \cite{Sato:2019yez, Moffat:2021dji}, a suppression of the unpolarized strange quark PDF relative to lighter quark PDFs was found to be preferred by the simultaneous extraction of spin-averaged PDFs and fragmentation functions when including the standard deep-inelastic scattering and semi-inclusive deep-inelastic scattering electromagnetic observables, semi-inclusive annihilation in $e^+e^-$ collisions, and lepton-pair production in $pp$ reactions. In \cite{Cammarota:2020qcw}, the  global analysis was performed for all existing single-spin asymmetries, providing the empirical demonstration that the flavor-dependent nucleon tensor-charges agree within uncertainties with those computed directly in LQCD.

\subsubsection{LQCD and experimental global analysis}
Another new direction that has recently begun to be explored is the inclusion of LQCD data as Bayesian priors that can help overcome the difficulty in deconvoluting QCFs from experimental data. An advantage of this is that the lattice data can in principle have access into domains of the QCFs that are inaccessible experimentally or are difficult to determine  in particular processes and kinematics.
In \cite{Lin:2017stx}, for example, the isovector tensor charge from LQCD was used as a Bayesian prior to extract the nucleon transversity distribution. Similarly, in \cite{Bringewatt:2020ixn} exploratory studies were carried out to include off-the-light-cone matrix elements in a QCD global analysis of spin-averaged and spin-dependent PDFs, demonstrating some level of success and/or tensions depending on the type of observables and the associated QCFs.
For other  studies involving LQCD, see  \cite{DelDebbio:2020rgv,Karpie:2019eiq} and the following subsection. The combination of  LQCD results and experimental data in the framework of Bayesian inference provide new avenues for addressing the challenges imposed by the inverse problem in QCFs and facilitate reliable comparisons between data and theory.

\subsection{Lattice QCD}

Lattice field theory is the theoretical framework for understanding the properties of strongly interacting matter. The fundamental theory of strong interactions is QCD, a quantum field theory that requires non-perturbative computations to address low-energy hadronic physics. Lattice QCD  provides both a rigorous definition of QCD as well as a powerful tool for numerical computations.  
The basic computational task in LQCD is a Monte Carlo evaluation of multi-dimensional integrals that results in the lattice regularization of QCD. 

Sophisticated and powerful QCD algorithms have been developed to take advantage of modern day supercomputers leading to many  important results aiding experimental efforts to understand the nature of strongly interacting matter. However, despite impressive achievements, computations very close to the continuum limit that are required to reduce systematic errors are still out of reach with today's resources. Machine learning methods offer a new avenue for improving current computations as well as allowing for studies of previously impossible questions.  
Typical LQCD calculations proceed in two stages. First, an ensemble of gauge field configurations is generated. Subsequently,  certain correlation functions of the fundamental fields are computed by averaging over the ensemble of gauge configurations.  Finally, analyses of these correlation functions lead to the desired physical observables. Machine learning  techniques are now applied to all these stages of LQCD computations promising to enhance and extend the current state of the art.

\subsubsection{The sign problem at non-zero density}
Systems at non-zero density (as in nuclear/neutron matter) or Minkowski time dynamics (parton distribution functions and transport coefficients) cannot be studied  with standard Monte Carlo methods due to  the   fermion sign problem.  Recently, it has been shown that by evaluating the relevant  path integral over a field manifold deformed into complex space, the sign problem can be alleviated or even eliminated~\cite{Cristoforetti:2012su,Cristoforetti:2014gsa,Fujii:2015vha,Fukushima:2015qza,DiRenzo:2015foa,Ulybyshev:2019fte}.  Up to now, the choice of manifolds has been guided by either impossibly expensive calculations or (human) insight into particular models. Machine learning methods have begun to be applied, both in supervised and unsupervised learning modes \cite{Alexandru:2017czx,Alexandru:2018ddf,Alexandru:2018fqp,Bursa:2018ykf,Ohnishi:2019ljc,Wynen:2020uzx} to discover the integration manifolds that alleviate the sign problem and in certain cases allow for calculations previously not possible. 
Refinement of these methods opens an exciting new avenue  for understanding QCD at non-zero density as well as understanding real-time dynamics of strongly interacting matter. 
At non-zero temperature the problem of reconstructing spectral functions from Euclidean correlators arises. Recently, both VAE~\cite{Chen:2021giw} and GP~\cite{Horak:2021syv} have been used to solve the associated inverse problem.

\subsubsection{Ensemble generation}
Generating gauge configuration ensembles very close to the continuum limit has also been proven a daunting computational task~\cite{Schaefer:2010hu}. Present-day computations with lattice spacing below 0.05 fm are severely limited due to critical slowing down, i.e., large auto-correlation times in the Markov Chain used to generate the gauge configuration ensemble. The idea of trivializing maps~\cite{Luscher:2009eq} was introduced several years ago as a possible solution to this problem. It relied on an analytically determined map that can be used to change variables in the path integral resulting in a trivial integration weight. In \cite{Luscher:2009eq}, it was shown  that such maps exist and can be constructed as solutions to a gradient flow equation whose generator can be set up perturbatively in the flow time. Unfortunately, practical constructions of such maps have not been carried out even for simple models~\cite{Engel:2011re}. A successful application of this program was done in the context of stochastic perturbation theory~\cite{Luscher:2014mka} which is used to perform high order perturbative computations in QCD. Machine learning methods provide a new approach to discovering these maps. In fact, the idea of remapping variables with complex probability distributions to trivially distributed variables is something that is commonly done in ML.  
In that sense, adaptation of these methods to the context of Markov Chain Monte Carlo (MCMC) for lattice field theory is an exciting idea with enormous potential for solving the critical slowing down problem in LQCD. 

Applications of ML methods to lattice field theory calculations have already emerged. The scalar $\phi^4$ theory in two dimensions is one of the first models studied, where GAN were used to reduce autocorrelation times~\cite{Pawlowski:2018qxs}. In a  work by~\cite{Albergo:2019eim,Medvidovic:2020vum}, it was shown that indeed normalizing flows constructed via ANN can be trained to effectively sample the configuration space of two-dimensional   scalar $\phi^4$ theory up to lattice sizes of $14^2$. In addition, \cite{Hackett:2021idh} used flow-based methods to sample from multimodal distributions. In \cite{Nicoli:2020njz},
thermodynamic properties of 
scalar $\phi^4$ theory  in two dimensions were studied using normalizing flows to generate field configurations confirming the efficacy of this approach. These works demonstrated that critical slowing down may be eliminated in MCMC for scalar $\phi^4$ theory.  However, further studies and improvements of the approach~\cite{DelDebbio:2021qwf} demonstrated that the training cost of the flow based models scales badly as the critical point is approached and the correlation length and lattice size grow, indicating that further refinements of the approach are needed for elimination of critical slowing down. 
The flow-based methods for Monte Carlo sampling have already been extended to gauge theories~\cite{Luo:2020stn,Kanwar:2020xzo,Boyda:2020hsi,Tomiya:2021ywc,Favoni:2020reg,Luo:2021ccm},  fermionic theories~\cite{Albergo:2021bna}, and  to theories with a sign problem~\cite{Lawrence:2021izu}.
It is interesting to note that an entirely different approach in using ML methods for optimizing MCMC is the so called L2HMC approach, see \cite{levy2018generalizing}, which has been recently applied to the $U(1)$ gauge theory with fermions in~\cite{Foreman:2021ixr}.
In this approach, ML methods are used to construct a new flow based map that replaces the Hamiltonian evolution in Hamiltonian Monte Carlo~\cite{PhysRevX.10.021020,Murphy2012,Bishop2006}. This is a promising idea that could have impact in realistic MCMC calculations in lattice field theory.

\subsubsection{Correlation function estimators}
Calculations of observables in LQCD require the computation of quark propagators in the background of a large number of gauge configurations. Quark propagators are computed by solving a linear system of equations with a large  and sparse coefficient matrix.  Machine learning methods are  used~\cite{Pederiva:2020} to obtain low precision solutions in order to construct approximate observables from which full precision results are obtained via the so called all-mode-averaging technique~\cite{Shintani:2014vja}, resulting in enormous increase in efficiency of the computation. Along similar lines, boosted decision trees were used as approximate estimators of two- and three-point correlation functions used in calculations of nucleon charges and of the phase acquired by the neutron mass with a small parity violation \cite{yoon2019}. The same methodology was also employed for correlators used in parton distribution function calculations~\cite{Zhang:2019qiq}. In addition, GP were employed in predicting the long distance behavior of matrix elements in PDF computations in LQCD~\cite{Alexandrou:2020tqq}. An interesting new idea was introduced in \cite{Detmold:2021ulb} where ML methods were used to find deformations of the fields in the path integral so that the variance of observables is reduced.
 
 \subsubsection{Miscellaneous}
 Finally there is a large body of work in which ML methods are used to understand properties of lattice field theories. These works include the prediction of lattice action parameters from field configurations~\cite{Shanahan:2018vcv,Blucher:2020mjt}, discovering the holographic geometry in AdS/CFT descriptions of field theories~\cite{Hashimoto:2018ftp,Hashimoto:2018bnb,Hashimoto:2019bih,Hashimoto:2020jug,You:2017guh}, and  understanding the nature of phase transitions~\cite{Bachtis:2020ajb,Wetzel:2017ooo,Blucher:2020mjt,Boyda:2020nfh,Chernodub:2020nip}. A discussion of ML applications to LQCD can also be found in a recent Snowmass report~\cite{SnowMass2021}.

\subsection{High-Energy Nuclear Theory}

At extremely high temperature or density, quarks and gluons are deconfined from nucleons to form a new state of nuclear matter -- the quark gluon plasma (QGP). In the early universe, this state of matter existed for about a few microseconds after the big bang and might also be at the core of some neutron stars or created during violent neutron star mergers.  The field of high energy nuclear physics aims to search for its formation, the phase transition between normal nuclear matter and QGP, locate the critical point(s) and determine the deconfinement temperature in high-energy heavy-ion collisions (HIC).  Many 
properties of QGP, such as transport coefficients  (e.g., shear and bulk viscosity, conductivity and jet transport coefficient) can be extracted from experimental data using soft and hard probes.

\subsubsection{Bayesian inference}

Since the lifetime of QGP in HIC  is quite short, about $10^{-23}$ seconds, experiments detect  the relics of QGP -- hadrons, photons, leptons and their momentum distributions and correlations. The properties of QGP can only be extracted indirectly through theoretical models. In practice, each observable in experimental data are entangled to many model parameters.
This entanglement hinders the determination of a specific model parameter. Even worse, sometimes different combinations of model parameters produce degenerate outputs when the data are projected into lower dimensions.

Two promising ML tools emerge to solve this seemingly ill-defined inverse problem. The first tool is  BO which tends to fit all model parameters at the same time using all available data. During this global fitting, if several groups of model parameters generate the same output, then the probability that the real model to be one of them decreases. 
On the contrary, if only one group of model parameters describes data, and changing parameters leads to large model-data differences, then the posterior distribution peaks at the optimal parameter value with small width -- indicating a small model uncertainty. 

Bayesian optimization is widely used in high energy nuclear physics to constrain the QCD equation of state \cite{Pratt:2015zsa}, the QGP transport coefficients \cite{Bernhard:2016tnd,Bernhard:2019bmu,Paquet:2020rxl}, fluctuation and correlation \cite{Yousefnia:2021cup}, the heavy quark diffusion coefficient \cite{Xu:2017obm}, jet transport coefficient \cite{JETSCAPE:2021ehl} and the jet energy loss distributions \cite{He:2018gks}. In many applications, other ML tools are used to assist the Bayesian analysis. For example, the relativistic fluid dynamic simulations of HIC are time consuming which prevents a fast MCMC random walk in the parameter space. Emulators based on the theory of GP are employed to approximate model outputs using efficient interpolations with much fewer design points in the parameter space. Since data obtained this way are redundant and correlated, PCA is used to compress data to lower dimensions.

The observables used in Bayesian inference are high-level features designed with personal experience. However, feature engineering is known to be incomplete, insufficient and sometimes misleading which may lose important correlations hidden in high-dimensional data. This is unavoidable due to the high-dimensional character of the experimental data. It is difficult to recognize from exotic low-level features the non-linearly correlated patterns that are unique and robust for determining a specific model parameter.

\subsubsection{Inversion problems with ML}

Deep neural networks are promising ML tools to tackle the difficult inverse problem in HIC. If low-dimensional projections of model outputs are degenerate in the seemingly ill-defined inverse problem, one would expect differences still exist in high dimensions or in non-linear correlations between different dimensions. The universal approximation theorem \cite{Cybenko1989,Hornik1991} ensures that ANN  have enough representation capability to map low-level features to some given model parameters in supervised learning. If information gets totally lost in the dynamical evolution of HIC  because of entropy production, the network can never succeed in building this map. If the information of one specific model parameter survives the dynamical evolution and exists in the final output of HIC, the network has a better chance to build this map. In this sense, if the network provides high accuracy prediction in a supervised learning, it indicates that the signal does encode in the complex final state output and the network helps to decode this information. On the other hand, if the network fails, it indicates that either the information on the physical signal gets totally lost, or the used network does not have enough representation power. For the later, it is still possible with a deeper, wider or more suitable network.

The type of  nuclear phase transition is an important input to relativistic hydrodynamic simulations of HIC. At high  energies, LQCD predicts that the transition between QGP and hadron resonance gas is a smooth crossover. At intermediate beam energies, it is conjectured to be a first order phase transition. Different phase transitions lead to different pressure gradients around the transition temperature which drives the QGP expansion (it also depends on the shear and bulk viscosity).  In this inverse problem, one can determine the phase transition type during the dynamical evolution using final state output. Supervised CNN \cite{Pang:2016vdc,Du:2019civ}, point cloud networks \cite{Steinheimer:2019iso}, and unsupervised AE \cite{Wang2020} are trained to identify the QCD phase transition types using the final state hadrons. It is verified that signals of the phase transition survive the dynamical evolution, and deep learning succeeds in decoding this information from the final hadron distribution. To avoid overfitting to given model parameters, different parameter combinations are used to form a diverse training dataset. Prediction difference analysis is used to interpret which region in the momentum space is most important for the networks to make their decisions.

\begin{figure}[hbtp]
    \includegraphics[width=0.98\linewidth]{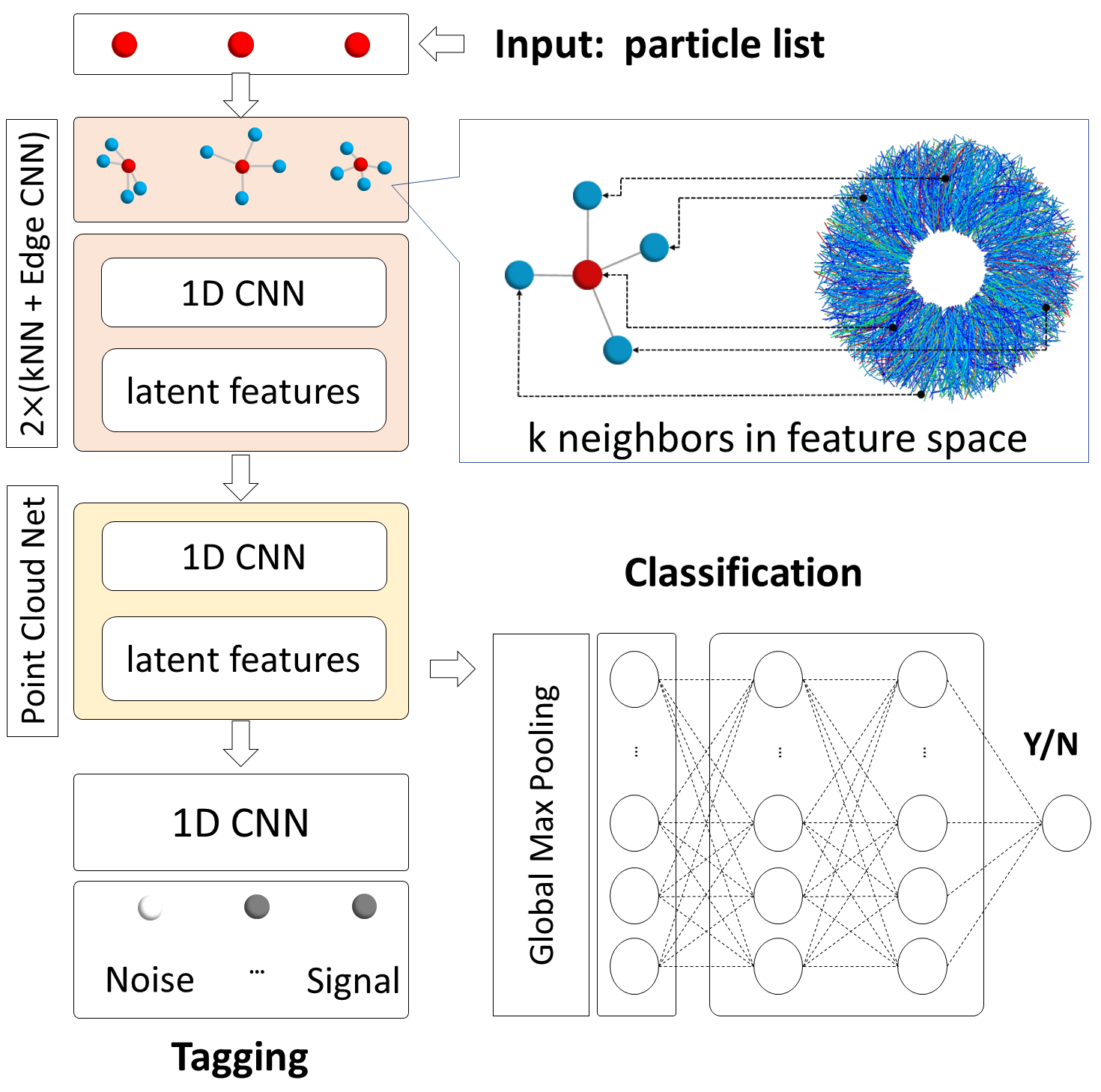}
    \caption{{\bf Dynamical edge-convolution neural network used for event classification and particle tagging in heavy ion collisions.}
    Figure taken from \cite{Huang:2021iux}.
    }
    \label{fig:edgeconv}
\end{figure}

Data produced in high-energy HIC are lists of particles with their four-momenta and quantum numbers. Early studies used histograms to convert this information into images which were required by two-dimensional CNN. It was found later that the point cloud network is suitable for this data structure \cite{Steinheimer:2019iso}. To enhance the representation power, a dynamical edge-convolution network followed by a point cloud net is used to identify self-similarity and critical fluctuations in HIC \cite{Huang:2021iux}. Figure~\ref{fig:edgeconv} shows the architecture of the dynamical edge-convolution network. The input to the network is a list of hadrons. The output has two branches: one for event classification and the other for particle tagging. The KNN algorithm finds the $k$-nearest neighbors of each particle in both momentum space and feature space. Repeating the KNN and edge-convolution blocks twice helps to find long-range multi-particle correlations that are the key to searching for critical fluctuations.

The impact parameter is the transverse distance between colliding nuclei whose precise determination helps many downstream tasks. It is not possible to measure impact parameters directly through experiment. Several different ML tools are used to determine the impact parameters of HIC, including shallow neural networks \cite{Bass:1996ez}, ANN \cite{Kuttan:2021zcu, Kuttan:2020kha}, SVM \cite{DeSanctis:2009zzb}, and EMB \cite{Mallick:2021wop,   Li2020,Li2021}. In this inverse problem, data from Monte Carlo simulations are used to map final state output to the impact parameter. The traditional method uses a single observable (the particle multiplicity) to determine the impact parameter. Machine learning methods using high dimensional data result in much smaller  uncertainties.

Another inverse problem associated with initial states is linked with the given nuclear structure, which in turn affects in many ways the final state outputs of HIC. In a prototype inverse problem, a 34-layer residual neural network is used to predict the deformation parameters of the involved nuclei \cite{Pang:2019aqb}. Using the simulation data, the network succeeds in extracting the magnitude of the nuclear deformation but fails to extract the sign. The failure denotes a degeneracy raised in the dynamical process of high energy collisions. In another inverse problem, Bayesian CNN are employed to identify the 3-$\alpha$ and 4-$\alpha$ structures in the colliding light nucleus, from the final output of simulated heavy ion collisions \cite{He:2021uko}. The overall classification accuracy reaches $95\%$ for $^{12}$C/$^{16}$O + $^{197}$Au collisions.

\subsubsection{Other applications of ML methods}

High energy jets lose energy when they traverse through  hot deconfined nuclear matter.  In the inverse problem,  CNN are employed to predict the energy loss ratio from final state hadrons inside a jet cone \cite{Du:2020pmp,Du:2021pqa}, which allows to study the jets based on the initial energy. 

The chiral magnetic effect  was expected to arise from possible parity violation of strong interactions. However, all previously proposed observables suffer from large background contamination in heavy ion collisions. Deep CNN are used to identify the charge separation associated with the chiral magnetic effect \cite{Zhao:2021yjo}. The network is robust to diverse conditions including different collision energies, centralities, and elliptic flow backgrounds.

The interaction between bottom and anti-bottom quarks in QGP is modeled as a heavy quark potential, whose variational function form is represented by  deep neural networks \cite{Shi:2021qri}. The inputs to the networks are the temperature of the QGP and the quark anti-quark separation $r$. The output is the heavy quark potential.  Solving the pertinent differential equations numerically with this potential gives mass spectra whose difference with LQCD calculations  defines the cost function. Optimizing this cost function gives the parameters of the model.

Finally, in Ref.~\cite{mroczek2022}, the authors use active learning \cite{cohen2018} to reduce sampling requirements for training classifiers in searches  for acceptable
EoS parameters.

\section{Experimental Methods}\label{Experiment}

The aim of this section is to give the reader an overview of recent progress and future research directions in ML approaches and methods applied to nuclear physics experiments. During the last few years, ML methods have been applied to the full chain of experimentation including the design of experiments, the acquisition of data, the processing chain of converting detector information into observables, and  physics analysis.

\subsection{Streaming Detector Readout}

In trigger-less or streaming readout data acquisition systems, detector data are read out in continuous parallel streams that are encoded with information about when and where the data were taken. This simplifies the readout as no custom trigger hardware and firmware is needed and is beneficial for, e.g., experiments that are limited by event-pileup or overlapping signals from different events. Streaming readout also gives opportunity to streamline workflows for online and offline data processing and allows to take advantage of ML approaches. 

The LHCb experiment at CERN has pioneered the idea of seamless data processing from the readout to the analysis, using software stages in early stages of the event selection. Part of this system is a custom boosted decision tree algorithm for the reconstruction of decay products of $b$-hadrons \cite{gligorov2013efficient,likhomanenko2015lhcb}. Progress in novel boosted decision tree algorithms allow improvements in the efficiency of decay classifications by up to 80\% for high-rate events. 

The CLAS12 experiment at Jefferson Lab tested a prototype streaming readout system successfully under beam conditions. The test was limited to the measurement of inclusive electroproduction of neutral pions in a lead-tungstate calorimeter and a hodoscope. An unsupervised hierarchical cluster algorithm was utilized in real-time with real data taken in streaming readout mode to combine the time, position, and energy information at the hit level, and associate each hit with a cluster membership and an outlier score \cite{Lawrence2021}. The implementation  \cite{mcinnes2017hdbscan} allows to successfully reject noise hits and to identify clusters for diverse topologies and large hit multiplicities.

\subsection{Reconstruction and Analysis}

\subsubsection{Charged particle tracking}

Deep learning approaches using neural networks have the advantage of a high level of flexibility and robustness with a minimum of assumptions about the data. This offers an effective solution to the challenges of charged particle tracking at high luminosity. At high luminosity, tracking suffers from track candidates that share hits (combinatorials). This results in hits wrongly identified as ``on-track'' and produces ghost tracks. 
In high luminosity environments, the largest fraction of CPU time in tracking in a traditional analysis is spent on setting up various Kalman filters \cite{kalmanfilter} at each measurement site. These are affine operations involving CPU and memory intensive matrix-matrix multiplications and matrix-vector multiplications. 
Therefore, the improvement in track seeding resulting from using ANN and deep learning methods, yields substantially faster track reconstruction speed.  In addition, the selection of the correct seed results in improved tracking efficiency \cite{guest2018,Gavalian2020}. Furthermore, noise rejection algorithms have an impact in selecting the right combinations of hits in seeding~\cite{Komiske2017}. 

One of the most common deep learning algorithms employed for tracking pattern recognition are CNN. The CNN features are generally representations of the detector geometry. Machine learning algorithms used for background rejection involve topological properties of tracks to isolate signal from background\cite{STAR:2020vfa}. Outlier detection methods are used to remove noise uncorrelated hits and to classify tracks including pile-up \cite{Ayyad:2018zzl}.

Large liquid scintillator detectors can hold a large target mass for neutrino detection, for example in solar and reactor neutrino experiments. Machine learning can benefit pure liquid scintillator detectors in event position and energy reconstruction~\cite{Qian:2021vnh}. Similarly, neutron detectors such as NeuLAND track both charged and neutral particles. Simple neural network architectures have been shown to efficiently assist in tracking \cite{Mayer2021}.

\subsubsection{Calorimetry}

The GlueX experiment at Jefferson Lab used ANN to reduce background in the GlueX forward calorimeter for the detection of photons produced in the decays of hadrons \cite{Barsotti2020}. Energy deposition characteristics in the calorimeter such as shapes, size, and distribution were employed to discriminate between signal and background, where the background mostly originates from hadronic interactions that can be difficult to distinguish from low energy photon interactions. The training was done on data using $\omega$-meson decays. The ANN-based algorithm showed to be a powerful tool to reconstruct neutral particles with high efficiency and to provide substantial background rejection capability.

\subsubsection{Particle identification}

 Particle identification (PID) is done with dedicated detectors capable of identifying certain particle types. 
For example, Cherenkov detectors are largely used in modern nuclear experiments for identifying charged particles like pions, kaons, and protons corresponding to a wide range in momentum. 
Cherenkov detectors are typically endowed with single photon detectors and the particle type can be recognized by classifying the corresponding detected hit pattern \cite{fanelli2020machine}.
DeepRICH \cite{fanelli2020deeprich} is a recently developed custom architecture that combines VAE, CNN and ANN. 
 The reconstruction performance is close to that of other established methods like FastDIRC \cite{hardin2016fastdirc} with fast reconstruction times due to its implementation on GPUs that allow for parallel processing of batches of particles during the inference phase.

In \cite{derkach2020cherenkov}, GAN have been used to simulate the Cherenkov detector response. This architecture predicts the multidimensional distribution of the likelihood for particle identification produced by FastDIRC by-passing low-level details. Here, GAN have been used in \cite{maevskiy2020fast} to have a fast and accurate simulation of Cherenkov detectors; these have been trained using real data samples collected by LHCb.    

The utilization of jets produced by the hadronization of a quark or gluon in heavy ion experiments or future electron ion colliders for nuclear physics such as EIC can be functional to a variety of fundamental topics \cite{page2020experimental}. Machine learning has been applied to design experimental observables that are sensitive to jet quenching and parton splitting \cite{Lai:2020byl,Lai:2021ckt}.
In \cite{chien2019probing}, a deep CNN  trained on jet images allowed the study of jet quenching utilizing quark and gluon jet substructures. Moreover, CNN  have been used to discriminate quark and gluon jets in \cite{komiske2017deep}.
Different deep architectures (CNN, Dense ANN, and RNN) have been studied in \cite{apolinario2021deep} for the classification of quenched jets, and in particular to discriminate between medium-like and vacuum-like jets.
Identification of jets as originating from light-flavor or heavy-flavor quarks is an important aspect in inferring the nature of the particles produced in high-energy collisions. 

Much progress has been made in recent years on heavy-flavor tagging, with custom architectures like DeepJet \cite{bols2020jet} and JetVLAD \cite{bielvcikova2021identifying}, and on strange jet tagging \cite{erdmann2020tagger, nakai2020strange}. 
Hadronic jets feature multiple tracks and extended energy deposition in both electromagnetic and hadronic calorimeters, and can represent a primary source of backgrounds for electrons. 
Recent studies based on CNN showed the advantage of using low level calorimeter data represented as images in identifying electrons \cite{collado2021learning}.
Electromagnetic showers have been classified using computer vision techniques that take advantage of lower level detector information  \cite{de2020electromagnetic}.  
Finally, BNN have been used for the pion, kaon, and proton identification with tests done on data generated for the BES II experiment \cite{ye2008applying}, combining multiple features from different detectors like drift chamber, time of flight, and shower counter.

\subsubsection{Event and signal classification}  

In collision-based experiments, events are often categorized by event type for analysis. Although selection algorithms differ across experiments, the selection is typically computationally expensive in traditional analyses. In high-luminosity experiments, real-time triggers are deployed to decide which events to store for analysis. For such algorithms, fast inference speed is required. In low-energy, low luminosity collisions, such as those at rare isotope facilities, event selection post-trigger is a computational bottleneck for big data detectors such as time projection chambers \cite{Bradt2017, nova2016}. 

A common task in scintillator detectors in low-energy experiments is to discriminate between the neutron  and $\gamma$ signals. Neural network analysis of pulse shapes have been shown to effectively discriminate between these signals \cite{Doucet2020}.

In post-experiment analyses, deep ANN and CNN \cite{Kuchera:2018djs, Gavalian2020, Solli2020} were used to classify events. Notably, seeding networks with pre-trained architectures designed for image classification  allowed for fast training when events can be structured similar to images. Additionally, hierarchical clustering \cite{DALITZ2019} is used for track finding in TPCs.

In low-energy neutrino experiments, ML techniques are typically used to differentiate different physics signal types or signals from backgrounds. One of the earliest uses of ANN to discern neutral-current from charged-current solar neutrino interactions in a heavy-water Cherenkov detector was performed by the Sudbury Neutrino Observatory experiment~\cite{Brice:1996pm}. 

Direct kinematic measurements of tritium $\beta$-decays have historically been the most sensitive to investigate the neutrino mass scale~\cite{Formaggio:2021nfz}. The next-generation tritium $\beta$-decay experiment Project8 plans to measure the $\beta$-electron energy spectrum via cyclotron radiation emission spectroscopy (CRES)~\cite{Monreal:2009za}. The geometry of the detector and its electromagnetic field configuration, as well as the dynamics of the $\beta$-electrons, would give rise to different variations of the CRES signals. The collaboration is developing machine learning techniques to analyze and improve the reconstruction of these CRES events~\cite{ref:p8_signal_classification}.

To answer the question of whether neutrinos are their own antiparticles, i.e., Majorana fermions, there are recently completed, operating, and planned experiments to search for neutrinoless double-beta ($0\nu\beta\beta$) decays in $^{76}$Ge~\cite{GERDA:2020xhi,Majorana:2019nbd,LEGEND:2021bnm}, $^{100}$Mo~\cite{CUPID:2019imh}, $^{130}$Te~\cite{CUORE:2014erp,SNO:2020fhu,Adams2022}, $^{136}$Xe~\cite{KamLAND-Zen:2016pfg,EXO-200:2019rkq,nEXO:2018ylp,NEXT:2013wsz}, and other isotopes. The observation of this lepton-number-violating decay mode, in which two electrons but no neutrinos are emitted, is evidence that neutrinos are Majorana fermions. As the current limit of this decay mode is on the order of 10$^{26}$~y, experiments --- whether they are large-scale liquid scintillator detectors, semiconductor ionization detectors, cryogenic bolometers, liquid cryogenic or high-pressure gaseous time projection chambers (TPC) --- are deploying machine learning techniques to distinguish and remove backgrounds in the signal search region. The ``single-site'' signature of a $0\nu\beta\beta$ decay signal is the simultaneous appearance from the same origin of two electrons whose energies add up to the decay's $Q$-value. Backgrounds tend to have ``multi-site'' characteristics; for example, external gamma rays that Compton-scatter at multiple locations in the detector's active volume. For comprehensive reviews of $0\nu\beta\beta$ decays, see \cite{Dolinski:2019nrj,Agostini2022}.

The use of high-purity germanium detectors (HPGe) enriched in $^{76}$Ge in $0\nu\beta\beta$-decay search, has a long history~\cite{Avignone:2019phg}. The GERDA experiment pioneered using ANN to identify single-site events in semi-coaxial high-purity germanium detectors~\cite{Agostini:2013jta,Agostini1445}. The use of ANN to differentiate single-site and multi-site events in HPGe was also developed in~\cite{Caldwell:2014joz,Jany:2021rqs,Holl:2019xtt}.

The NEXT experiment exploits the topological difference between $0\nu\beta\beta$-decay event signal and backgrounds in its $^{136}$Xe high-pressure gas TPC using deep neural networks~\cite{Renner_2017}, and specifically  CNN~\cite{Kekic:2020cne} for the identification of event topology similar to that of a signal event. The ML tools developed by NEXT have also been adapted in the conceptual design of a $^{82}$SeF$_6$ TPC for $0\nu\beta\beta$-decay study~\cite{Nygren_2018}. 

In $0\nu\beta\beta$-decay experiments, an ample amount of the target isotope can be loaded in the scintillator, but the detector energy resolution is typically worse than other types of detectors. The KamLAND-ZEN experiment developed CNN and RNN to identify $^{10}$C from cosmic-ray spallation in the liquid scintillator loaded with $^{136}$Xe~\cite{Li:2018rzw,Rojo:2018qdd}. Efforts in developing liquid scintillators that allow the separation of the Cherenkov light and the scintillation will facilitate ML to perform signal-background differentiation in $0\nu\beta\beta$-decay experiments or other multi-purpose detectors~\cite{Askins:2019oqj,gruszko_julieta_2018_1300693,fraker_suzannah_2018_1300914}. Large water-Cherenkov detectors could also benefit from gadolinium loading to enhance their neutron detection capability for supernova detection and searches of the diffuse supernova neutrino background. The use of ML for background discrimination in such a detector has been studied~\cite{Maksimovic:2021dmz}.  

\subsubsection{Event reconstruction}

For the  precise knowledge of the kinematic variables of the deep inelastic scattering process, various reconstruction methods are combined in collider experiments. Each method is using partial information from the scattered lepton and/or the hadronic final state of  deep inelastic scattering and has its own limitations. Recently, it has been shown for the H1 and ZEUS collider experiments as well as for simulations of a possible EIC detector that deep learning techniques to reconstruct the kinematic variables can serve as a rigorous method to combine and outperform existing reconstruction methods~\cite{Diefenthaler:2021rdj,Arratia:2021tsq}. 

\subsubsection{Spectroscopy}

Gamma-ray spectra are used, among other things, for isotope identification and fundamental nuclear structure studies. Timing resolution in high-purity germanium detectors have been optimized by fully-connected CNN architectures \cite{Gladen2020}. Deep, fully-connected neural network architectures are shown to successfully identify isotopes
\cite{Kamuda2017, Abdalaal1997, Medhatc2012, Jhung2020} and fit peaks \cite{Abdel2002} in gamma spectra. Machine learning has also been shown to estimate activity levels in spectra from gamma-emitting samples \cite{Abdalaal1997, Kamuda2018, Vigneron1996}. 
Convolutional neural networks have demonstrated robustness to spectra with unidentified background channels and calibration drifts in the detectors \cite{Kamuda2020}. In addition, charged particle detection is routinely used for spectroscopy. For example, \cite{Bailey2021} uses ML to analyze signals from double-sided silicon strip detectors to determine $\alpha$-clustering.

\begin{figure*}[!htb]
\includegraphics[width=0.9\linewidth]{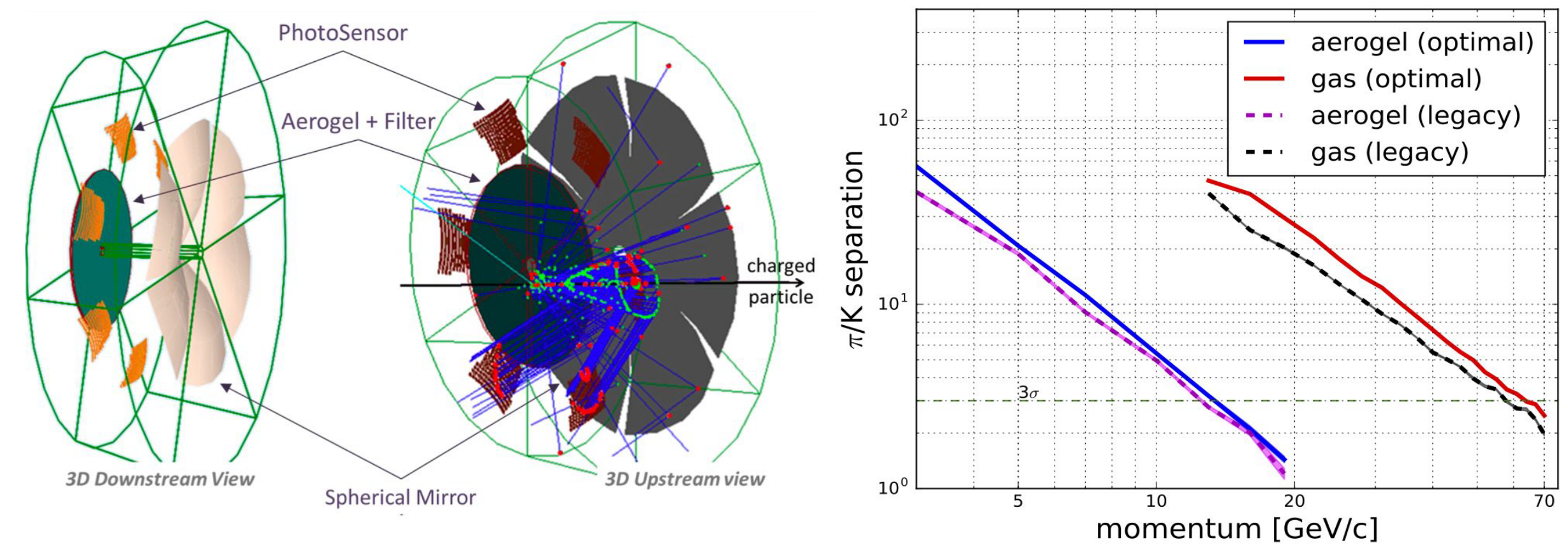}
\caption{
\textbf{(left) \textsc{Geant4} model of the d-RICH.}
The full 3D downstream (left) and upstream (right) views of the d-RICH detector. 
The BO strategy involves tuning eight main design parameters characterizing the geometry and the optical properties. 
{\bf (right) $\pi / K$ separation power.} 
$\pi / K$ separation as number of $\sigma$ as a function of the charged particle momentum. The improvement in the separation power with the approach is compared to the legacy baseline design. The curves are drawn with 68\% confidence interval bands which are barely visible in the log plot.
Taken from \cite{cisbani2020ai}.
}
\label{fig:posterior}
\end{figure*}
\subsection{Experimental Design}\label{Design}
\subsubsection{Design for detector systems}

Physics and detector simulations are critical for both the initial design and the optimization of complex sub-detector systems in nuclear physics experiments.
These systems are usually characterized by multiple parameters capable of tuning, e.g., the mechanics, the geometry, and the optics of each component. 

Traditionally the global design is studied and characterized after the subsystem prototypes are ready. In the sub-detector design phase, constraints from the baseline full detector are taken into account. 
A well-known phenomenon observed in optimization problems with high-dimensional spaces is the so-called ``curse of dimensionality'' \cite{houle2010can}, corresponding to a combinatorial explosion of possible values to search.  
Indeed, the detector design optimization can be a large combinatorial problem characterized by accurate and computationally expensive simulations performed in {\sc Geant4} \cite{agostinelli2003geant4}. 
In this context, ML offers different optimization strategies, spanning from reinforcement learning \cite{sutton2018reinforcement} to evolutionary algorithms \cite{deb2001multi}.
These new approaches can potentially ``\textit{revolutionize the way experimental nuclear physics is currently done}'' \cite{stevens2020ai}. 
Among these approaches, BO \cite{jones1998efficient, snoek2012practical} has gained popularity in  detector design because it offers a derivative-free principled approach to global optimization of noisy and computationally expensive black-box functions. 
An automated,  highly-parallelized,  and self-consistent procedure has been developed and tested for a dual-radiator Ring Imaging Cherenkov (d-RICH) design \cite{cisbani2020ai}, which has been considered as a case study.
Eight main design parameters have been considered to improve the particle identification (PID) performance of the d-RICH detector. Examples of design parameters  are the refractive index and thickness of the aerogel radiator, the focusing mirror radius, its longitudinal and radial positions, and the three-dimensional  shifts of the photon sensor tiles with respect to the mirror center on a spherical surface. 

Gaussian processes have been used for regression, and a surrogate model has been reconstructed as shown in Fig.~\ref{fig:posterior}.
These studies not only resulted in a statistically significant improvement in the PID performance compared to an existing baseline design, but they also shed light on the relevance of different features of the detector for the overall performance.  

Remarkably, the future Electron Ion Collider (EIC) is looking at systematically exploiting AI-based optimization strategies during the design and R\&D phase of the EIC detector \cite{khalek2021science}.

A pipeline based on machine learning for the LHCb electromagnetic calorimeter R\&D optimization has been proposed \cite{boldyrev2020ml,boldyrev2021machine} to evaluate its operational characteristics and determine an optimal configuration with a gain in computational time. 
Different cell sizes have been considered to characterize the calorimeter granularity. 
Signal energy deposits in the calorimeter have been studied for different background conditions by changing the number of primary vertices. Energy and spatial reconstruction were based on gradient-boosted decision trees \cite{xgboost2016}, with the regressor trained to minimize the difference between reconstructed and generated observables.
 Machine Learning methods inside this pipeline are used to fine-tune the parameters of both simulations and reconstructions. 

The application of tools based on deep neural networks and modern automatic differentiation techniques to implement a full modeling of an experimental design, in order to achieve an end-to-end optimization is described in \cite{baydin2021toward}.

\subsubsection{Interface with Theory}

\begin{figure}[!htb]
\includegraphics[width=1.0\linewidth]{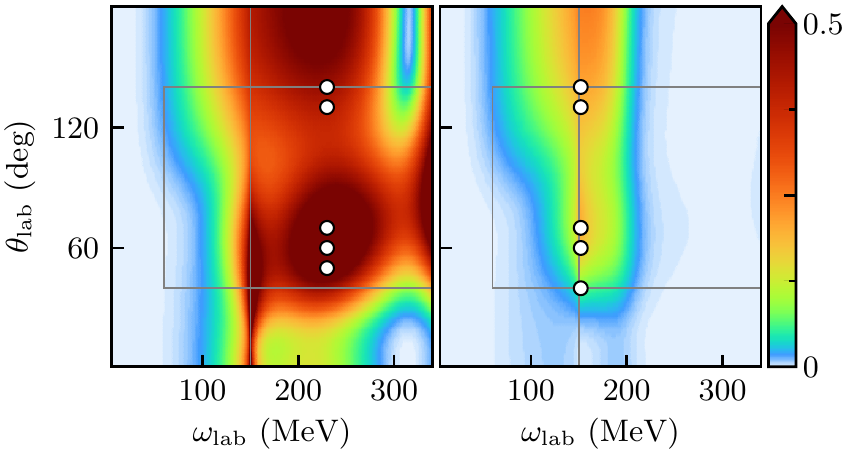}
\caption{\label{utility} {\bf BED of a proton Compton scattering experiment.}   Shown is the expected utility of proton differential cross section measurements defined as the gain in Shannon information based on the experiment. Chiral EFT has been used to predict the functional form of the scattering amplitude. The left panel does not include EFT truncation estimates, whereas the right panel includes the EFT uncertainty. The interior box marks the experimentally accessible regime. The vertical line marks the cusp at the pion-production threshold. The white circles show the optimal five-point design kinematics. Adopted from \cite{Melendez2021}.}
\end{figure}

Bayesian experimental design provides a general framework to maximize the success of an experiment  based on the best information available on the existing data, experimental conditions (including the amount of beam time available, experimental setup, budgetary constraints), and theoretical models used to process and interpret the data. The goal of  BED is to maximize the expected utility of the outcome. Formally this is done by  introducing the utility function that is designed by taking into account all costs and benefits
 \cite{Phillips2021}. Recent applications of BED can be found in  \cite{Melendez2021,Giuliani2021}. Figure~\ref{utility} shows the utility function for proton Compton scattering experiment interpreted using chiral EFT. 
As discussed in \cite{Melendez2021}, the effect of including EFT truncation errors is significant; it shifts the region of optimal utility to lower energies. The authors of Ref.~\cite{Giuliani2021} used 
the transfer function formalism in BED of experiments, aiming at the extraction of neutron densities.

\section{Accelerator Science and Operations}\label{Accelerators}

Research in modern ML techniques for accelerators is relatively new, however, it is a very active cross-cutting effort between data scientists, computer scientists, control experts, and accelerator physicists. 
Machine learning applications are poised to play an important role in accelerator facilities by providing data-driven digital models/twins for anomaly detection and prognostication, design optimization tools, and real time operational control/tuning.
There have been previous efforts to document opportunities \cite{edelen2018opportunities} and advanced control \cite{scheinker2020advanced} in ML for accelerators.
The following is an overview of some of these activities and new advancements since those documents.

\subsection{ML-based surrogate models for accelerator models}
Particle accelerators are complex non-linear systems that require sophisticated simulation software to capture this dynamic. Due to their complexity and evolving conditions, it is natural to explore modern data science techniques to provide surrogate models and/or a fully realized digital twin \cite{Grieves2017,DT9103025}. 
By developing digital models, researchers are able to conduct exploratory research without impacting the physical system. 
The use of ML to develop these digital models provides the ability to capture non-linear complex dynamics by using techniques such as GP Regression~\cite{NIPS1995_7cce53cf}, Quantile Regression Models~\cite{QR} and RNN~\cite{RNN}. These techniques can also be used to predict the future condition of the accelerator system and study anomalies, and to forecast component fatigue and failures. 

For simulation studies, surrogate models are the most popular examples of ML methods being used in the accelerator community to map between various accelerator parameters and beam properties at speeds that are orders of magnitude faster than computationally expensive physics models and for optimization studies~\cite{li2018genetic_accel_ML,edelen2020machine,kranjvcevic2021multiobjective_accel_ML,emma2018machine_accel_ML,hanuka2021accurate_accel_ML,zhu2021deep_accel_ML,scheinker2020adaptive_accel_ML}. Additionally, purely data-driven approaches have been explored to model accelerator components, such as the Fermi National Accelerator Laboratory (FNAL) booster system~\cite{john2021realtime, Kafkes:2021jse}.

\subsection{Anomaly detection and classification}
There is a growing amount of research using ML to improve accelerator operations by detecting anomalies and classifying them.
These studies explore several techniques that identify and elucidate the source of the anomalies and ultimately prevent them altogether in the future. 
Although it can be relatively easy to avoid these failures when they are actually happening it would be far better to avoid them all together. 
Unfortunately, forecasting these failures can be difficult to predict accurately with an acceptable false positive rate. 
Predictive diagnostics are important for improving operational efficiency, however, they can also serve as actionable precursors for control systems. 
There is a growing body of research in this domain, as such we present a select few recent results.

Machine learning models were used to classify  cavity faults in the Continuous Electron Beam Accelerator Facility~\cite{PhysRevAccelBeams.23.114601, solopova2019}. 
Traditionally, accelerator operators are able to analyze time series data, identify which cavities faulted and then classify the fault type, however, manually labeling the data is laborious and time consuming. 
In order to accelerate the process, \cite{PhysRevAccelBeams.23.114601} developed ML models to identify the faulty cavity and classification fault type. 
The results showed that the cavity identification and fault classification models have an accuracy of $84.9\%$ and $78.2\%$, respectively.

In \cite{2020-Rescic-AccelFailure}
 several machine learning techniques were used to predict machine failures at the Spallation Neutron Source  facility of Oak Ridge National Laboratory using beam current measurements before the faults actually occurred. 
They evaluated these pulses using a common set of ML-based classification techniques and showed that they can identify accelerator failures prior to actually failing with almost $80\%$ accuracy. The results were further improved by tuning the classifier parameters and using pulse properties for refining datasets leading to nearly $92\%$ accuracy in classification of bad pulses. 

\cite{2021-Li-AccelTimeSeries} examine the potential to predict interlocks (reflecting beam shutoff) using multiple measurements along the accelerator. 
Recurrence-plots based CNN (RPCNN) were used to convert the measurements into  time recurrence plots that are used as input to CNN for classification. Comparisons with EMB methods like random forests indicate that the performance of the random forests and RPCNN are comparable, however, the RPCNN is more successful at identifying anomalies that evolve over time. 

A novel uncertainty-aware Siamese model to predict upcoming faults~\cite{blokland2021uncertainty} was developed that combines the use of uncertainty quantification and a deep Siamese architecture to predict the similarity between beam pulses and provide an uncertainty that includes out of domain errors. The results show that this model outperforms previous results by four times in operational regions of interest.
Additionally, the model performed better than previous results even for anomalies it had not seen before.  
The inclusion of uncertainty quantification is an important step towards having robust and safe machine learning models for operational facilities.

\subsection{Control optimization}
Similar to the efforts in anomaly detection and classification, the field of accelerator optimization has recently seen advancements by applying techniques such as BO, genetic algorithms, particle swarm, and reinforcement learning.  
By utilizing online data from existing fast diagnostics such as beam position monitors, beam loss monitors, and radiofrequency cavity signals, an effort has been made to develop ML-based controllers. Recent demonstrations include BO, RL, and GP for accelerator tuning \cite{shalloo2020automation,PhysRevLett.124.124801,hao2019reconstruction,li2019bayesian,mcintire2016bayesian,Miskovich2021, WangNeurIPS2021, Roussel2021} and polynomial chaos expansion-based surrogate models for uncertainty quantification \cite{adelmann2019nonintrusive}. Particle Swarm techniques have been evaluated to optimize the tuning of aperiodic ion transport lines and for advanced particle separators \cite{Amthor:2018}. The use of multi-objective genetic algorithm-based (MOGA) optimization was used to minimize competing objectives steering operations of CEBAF  linacs \cite{PhysRevSTAB.17.101003} and showed that the dynamic heat load can be reduced by over $20\%$ while keep the same trip rate.
Additional studies using MOGA for accelerator optimization include \cite{PhysRevAccelBeams.21.054601, PhysRevAccelBeams.22.054602}.  Reinforcement learning tools have also been developed to optimize various elements of the accelerator system \cite{john2021realtime, o2020policy,bruchon2020basic,kain2020sample,hirlaender2020model,PhysRevAccelBeams.22.014601}. Additionally, there is new research on transferring the RL policy models to FPGA to provide a low latency control response time~\cite{john2021realtime}.
\color{black}
The accuracy of the ML methods for accelerators described above quickly degrades for systems that change with time, for which previously collected training data are no longer accurate. A major open problem faced by the ML community is the challenge of developing ML tools for complex systems where the underlying dynamics is evolving \cite{shimodaira2000improving,sugiyama2012machine,kurle2019continual,dramsch2021complex}. For systems that change very slowly with time and for which gathering large amounts of new data are feasible without interrupting operations, it is possible to utilize transfer learning techniques \cite{Goodfellow2016}. 

The most common transfer learning technique is to update or partially re-train the model by using new data \cite{calandra2012learning,koesdwiady2018non,kurle2019continual}. Another transfer learning approach is domain transform in which  smaller neural networks are developed using experimental data and used as the input layer of trained ANN \cite{zeiler2010deconvolutional}. These transfer learning techniques can be applied to GP-based algorithms in which the prior and parameter correlations are first estimated using simulation studies and then fine-tuned with experimental data. Such transfer learning techniques have been demonstrated on a wide range of systems including cross-modal implementations \cite{castrejon2016learning} and for electron back-scatter diffraction \cite{shen2019convolutional}.

\subsection{Adaptive ML for non-stationary systems}
For accelerators, new data can be acquired in many cases  in real-time and used to quickly update or re-train specific layers of neural networks. However, in other cases  repetitive re-training is not always an option. For example, if beam or accelerator characteristics change significantly diagnostics such as quadrupole magnet scan-based emittance measurements or wire-scan beam profile measurements can be time-consuming procedure that can interrupt operations. For quickly time-varying systems adaptive feedback techniques exist which are model-independent and automatically compensate for non-modeled disturbances and changes. Novel adaptive feedback algorithms have been developed to tune large groups of parameters based on noisy scalar measurements with analytic proofs of convergence and analytically known guarantees on parameter update rates,  making them especially well-suited for particle accelerator problems \cite{scheinker2013model,scheinker2017model}. Adaptive methods have been applied in real time to changing accelerators to maximize FEL output power \cite{scheinker2019model}, for real-time online multi-objective optimization \cite{scheinker2020online}, for non-invasive diagnostics \cite{scheinker2015adaptive}, and for online RL to learn optimal feedback control policies directly from data \cite{scheinker2021extremum}.  A limitation of local adaptive methods is getting stuck in local minima in high dimensional parameter spaces. 

Adaptive ML  and continuous learning is an area of active research combining the robustness of model-independent algorithms with the global properties of ML tools such as CNN. The latter have been combined with adaptive feedback for fast automatic longitudinal phase space control of time-varying electron beams \cite{scheinker2018demonstration}. 
Such adaptive ML tools have the potential to enable truly autonomous accelerator controls and diagnostics that  automatically respond to non-modeled changes and disturbances in real-time and thereby keep the accelerator performance (beam energy and energy spread, beam loss, phase space quality, etc.) at a global optimal.

\section{Nuclear Data}\label{Data}

The ``nuclear data pipeline'' represents the interconnected steps wherein data are compiled, evaluated, processed, and validated for end-user applications \cite{Romano2020}. Evaluations, the most labor-intensive step in the pipeline, provide a recommended ``best'' data value \cite{Capote2010} by folding together new measurements, previous measurements, and nuclear model predictions. In some cases, it can take several years for data to pass fully through the pipeline, limited by manpower and, in some cases, outdated methodology. Machine learning has the potential to significantly improve each step of the pipeline (compilation, evaluation, processing, validation, dissemination), which may notably reduce the time lag from data measurement to incorporation into standardized databases -- like the Evaluated Nuclear Data File (ENDF) \cite{Brown2018} -- that are  critical for basic and applied research. Machine learning can also facilitate the creation of surrogate models or emulators that may improve the extraction of physics information from data measurements, as well as improve the predictability of evaluation models. 

\subsection{ Overhauling the Nuclear Data Pipeline}
Creating a new ENDF release is a time-consuming and complex process whereby the latest experimental reaction measurements, theoretical reaction calculations, theory-generated reference parameter sets, and integral and benchmark experiments are combined and adjusted in an iterative manner for optimum consistency. An effort is underway \cite{Schnabel2021,Schnabel2020} to build software ``containers'' to hold all of these components, as well as the current evaluations of individual reactions. By appropriately nesting and interlinking these containers to mimic portions of the nuclear data pipeline (e.g., cross section measurements and models are linked together to produce an evaluation of one reaction, which is linked to other reactions to produce an evaluated library), they can be treated as interlinked nodes in a Bayesian network (BN). Then, GP can be used to automatically and self-consistently update the BN components -- including the output of a new ENDF reaction data library -- when any of the components are updated. The test cases to date utilize a sparse GP construction integrated into the BN framework to enable modeling of energy-dependent cross sections, physics model deficiencies, and energy-dependent systematic experimental errors. This would be a completely new approach to update ENDF, and one that could be quickly adapted to incorporate new benchmark experiment types, new theory codes, and new techniques to evaluate individual reactions. When completed, this approach could be modified for use in other fields that closely interlink experiment, theory, and benchmarks. 

\subsection{ Improving Compilations and Evaluations}

One utilization of ML to improve evaluations is through a robust identification of outlying data points and problematic datasets. A recent example \cite{Neudecker2020} was the identification of a problematic $^{19}$F neutron inelastic cross section in the ENDF database, via the use of EMB like  random forest regression combined with a SHapley Additive exPlanations (SHAP) feature importance metric \cite{lundberg2017} of predicted effective neutron multiplication factor $k_{eff}$ of critical assemblies. The ML approach identified this problematic cross section that was missed with traditional validation methods; this technique may be applicable for validating other data libraries against particular features of an end-user application. Another recent study \cite{Whewell2020} focused on utilizing a variety of ML approaches, including SVM, LR, and EMB, to find relationships between outlying measurements and underlying details of the experiments used in an evaluation (e.g., backgrounds, sample backing and impurities, detector type, incident beam type). Such an approach, which provides a robust method to reject outlying measurements and guide future experiments, was used in this specific case to improve the evaluation of the $^{239}$ Pu(n,f) reaction. Random forests were also used \cite{Neudecker2021} to combine differential data (differential cross sections),  application-specific integral data ($k_{eff}$ and neutron-leakage spectra from pulsed-sphere experiments), and nuclear theory to isolate likely root causes of disagreements between integral data and predictions using differential data, to select preferred differential datasets that give better agreement with integral measurements, and to specify which differential or theory developments can better reproduce integral results.

Machine learning can be used to provide quantified interpolations and extrapolations of nuclear data, see discussion on ML for data mining in  Sec.~\ref{LENT}. A good example is offered by the study of Giant Dipole Resonances (GDRs) \cite{Bai2021}. ANNs were first used to classify nuclei into groups with one or two GDR peaks, then two multi-task learning neural nets \cite{zhang2021survey} were used to learn both GDR energies and widths. The multi-task learning approach enables multiple related tasks to be learned simultaneously in a way that optimally utilizes the data for each task. As a result, the accuracy of predictions for measured nuclei was doubled compared to other approaches, and then the unmeasured properties of GDRs of  nuclei near the $\beta$-stability line were predicted. Another study \cite{osti_1823638} used ten different ML classifiers with a SHAP metric to automate and correct the assignment of spins to neutron resonances, a critical input to BO used in evaluations.

As discussed earlier in Sec.~\ref{LENT}, UQ in evaluations have also been improved by learning discrepancies from existing theoretical models. Moreover, ML can be used to design experiments that address nuclear data gaps for particular applications (e.g., criticality experiments \cite{Michaud2019}). The new data obtained can be essential for validation efforts in the pipeline which can provide critical adjustments of evaluations.

Machine learning is also being used to help extract knowledge from published documents. Here, CNN combined with edge detection techniques, are beginning to automate the extraction of data (tables, plots, numbers) from publications \cite{Soto2019}. Natural language processing (NLP), widely used to process text, is an ML-enhanced textual analytics approach that is now a central ingredient of efforts to revamp numerous standardized databases; this includes extracting keywords from documents as needed for the Nuclear Science References bibliographic database \cite{Pritychenko2011}. Next steps in NLP utilization in nuclear data may focus on the extraction of meaning (i.e., semantics) from documents. In this approach, data and concepts could be characterized as  ``high value'', and certain theoretical and experimental investigations could be recommended, based on latent knowledge in the literature \cite{Tshitoyan2019}.

\subsection{ Building Emulators and Surrogate Models}

Phenomenological models form the foundation of many evaluation approaches, because they are computationally inexpensive and often require few input parameters. As a result, the power to extrapolate evaluations to unmeasured nuclei, to perform self-consistency checks across the nuclear chart, and to accurately capture the correlations between different nuclear properties is limited. Machine learning has the potential to dramatically improve this, by creating emulators or surrogate models that require similar computational resources as phenomenological models but capture the physics of full models. A recent example from reactor physics \cite{RADAIDEH2020287} may provide a blueprint for similar advances in nuclear data. Here, deep learning ANN were used to perform UQ, sensitivity analyses, and uncertainty propagation for nuclear reactors by replacing high-fidelity reactor simulations with surrogate models.

\section{Summary and perspectives}

In recent years, ML techniques  have  gained considerable traction in scientific discovery. In particular, applications and techniques for so-called fast ML, i.e.,  high-performance ML methods applied to real time experimental data processing, hold great promise for enhancing scientific discoveries in many different disciplines \cite{deiana2021}. 
These developments cover a broad mix of rapidly evolving  fields, from the development of ML techniques to computer and hardware architectures. For a field like nuclear physics, which covers a huge range of energy and length scales, spanning from the smallest constituents of matter to the physics of dense astronomical objects such as neutron stars, AI and ML techniques offer possibilities for new discoveries and deeper insights. 

This Colloquium summarizes present and planned applications of ML techniques in experimental and theoretical nuclear physics research. The vast range of scales is also reflected in new and planned nuclear physics facilities worldwide, where opportunities to incorporate ML methods are expected to play an important role in  the justification and  design of experiments, and during the operations. In our overview, we have presented several recent experimental developments, including detector control, experimental design simulations, and accelerator operations. Furthermore, ML techniques play a central role in theoretical nuclear physics and a growing role in the evaluation of nuclear data. Nuclear theory, in particular, has seen an explosion in the application of ML methods in the last few years.
It is to be noted, however, that this Colloquium presents just a snapshot of ML in nuclear physics as of today. 
New ML approaches are being introduced continuously at a pace difficult to keep up with.
We expect thus that over the next decades ML will play a significant role in leveraging technologies that are at the frontier of computational science and data science.

Due to the broad range of scales, nuclear physics is a field where the dimensions of the problems studied  quickly exceed the capabilities of traditional computational methods. Machine learning techniques offer  promising paths to dimension reductions. Traditionally, many of the standard ML methods  focus  on making predictions and finding correlations in the datasets. 
However, as presented in this overview,  to quantify errors and find causations requires also the possibility of being able to determine models for probability distributions. Both in experiment and theory there are clear indications that  statistical learning methods offer new perspectives for future research directions. Research  in statistical learning techniques for both supervised and  unsupervised learning, combined  with for example fast ML methods and similar developments, have the potential to change the landscape of nuclear physics.

\begin{acknowledgments}
The work of AB, MD, NS, MS, and VZ is supported by the U.S. Department of Energy, Office of Science, Office of Nuclear Physics contract DE-AC05-06OR23177, under which Jefferson Science Associates, LLC operates the Thomas Jefferson National Accelerator Facility. NS is also supported by the U.S. Department of Energy, Office of Science, Office of Nuclear Physics in the Early Career Program. 
The work of CF is supported by the U.S. Department of Energy, Office of Science, Office of Nuclear Physics under grant No. DE-SC0019999.
The work of MHJ is supported by the U.S. Department of Energy, Office of Science, office of Nuclear Physics under grant No. DE-SC0021152 and U.S. National Science Foundation Grants No. PHY-1404159 and PHY-2013047. 
TH is supported by NSF grant PHY-2012430.
The work of MPK is supported by the U.S. National Science Foundation under Grant No. PHY-2012865. 
The work of DL is partially supported by the U.S. Department of Energy (DE-SC0013365 and DE-SC0021152) and the NUCLEI SciDAC-4 collaboration (DE-SC0018083). 
The work of WN is partially supported by the U.S. Department of Energy, Office of Science, Office of Nuclear Physics under Award Nos. DE-SC0013365 and  DE-SC0018083 (NUCLEI SciDAC-4 collaboration),  and by the National Science Foundation CSSI program under award number 2004601 (BAND collaboration). 
KO is supported by Jefferson Science Associates, LLC under U.S. DOE Contract \#DE-AC05-06OR23177, by U.S. DOE grant \#DE-FG02-04ER41302, and by the Center for Nuclear Femtography grants \#C2-2020-FEMT-006, \#C2019-FEMT-002-05.  The work of AP and XW is supported by Lawrence Berkeley National Laboratory under the U.S. Department of Energy Federal Prime Agreement DE-AC02-05CH11231.
The work of AS is supported by U.S. Department of Energy (89233218CNA000001) and by the Los Alamos National Laboratory LDRD Program (Project No. 20210595DR). The work of MSS is supported by the U.S. Department of Energy, Office of Science, Office of Nuclear Physics under grant No. DE-AC05-00OR22725.
The work of LGP is supported by the NSFC under grant Nos.~11861131009 and 12075098.
 
\end{acknowledgments}

\bibliography{References} 
\end{document}